\documentclass[aps,prc,superscriptaddress,floats,floatfix,reprint,reprintnumbers]{revtex4}
\usepackage{blindtext}
\pdfoutput=1
\usepackage{epsfig}
\usepackage{graphicx}
\usepackage{dcolumn}
\usepackage{bm}
\usepackage{booktabs}
\usepackage{color}
\usepackage{mathrsfs}
\usepackage{ulem}
\usepackage{epstopdf}
\graphicspath{{plots/}}
\usepackage{verbatim}
\usepackage{indentfirst}
\usepackage{float}
\usepackage[colorlinks,
           bookmarks=true,
           linkcolor=blue,
           urlcolor=blue,
            anchorcolor=black,
            citecolor=blue]{hyperref}
\usepackage[bf,raggedright,indentafter]{titlesec}
\usepackage{rotating}
\usepackage{booktabs}
\usepackage{tabularx}
\usepackage{makecell}
\usepackage{hypcap}
\usepackage{subfigure}%
\usepackage{tabularx}

\usepackage{array}

\begin{document}
\hspace{3cm}
    \title{\Large Study of the axial-vector and tensor resonant contributions to the $D \to VP\ell^+\nu_\ell$  decays based on  SU(3) flavor analysis}
\author{Yi Qiao,~~~Yue-Xin Liu,~~~Yuan-Guo Xu,~~~Ru-Min Wang$^{\natural}$\\
{\scriptsize College of Physics and Communication Electronics, Jiangxi Normal University, Nanchang, Jiangxi 330022, China}\\
$^\natural${\scriptsize Corresponding author.~~~~Email:ruminwang@sina.com}
 }

\vspace{3cm}

\begin{abstract}
Semileptonic three-body $D \to M\ell^+\nu_\ell$ decays,  non-leptonic $M \to VP$ decays, and semileptonic four-body $D \to M(M \to VP)\ell^+\nu_\ell$ decays are analyzed using the SU(3) flavor symmetry/breaking approach, where $\ell=e/\mu$,  $M=A/T$, and $A/T/V/P$ denote  the axial-vector/tensor/vector/pseudoscalar mesons, respectively.
In terms of SU(3) flavor symmetry/breaking, the decay amplitudes  of the $D \to M \ell^+ \nu_\ell$ decays and the vertex coefficients of the $M \to VP$ decays are related.  The relevant  non-perturbative parameters of the $D \to A\ell^+\nu_\ell$,  $A \to VP$ and  $T \to VP$ decays are  constrained  by the present experimental  data, and the non-perturbative parameters of $D \to T\ell^+\nu_\ell$ decays are taken from  the results in the light-front quark model since no experimental data are available at present. The branching ratios   of the $D \to M \ell^+ \nu_\ell $, $M \to VP$, and $D\to M(M\to VP)\ell^+\nu_\ell$ decays are then predicted. We find that
some processes receive both tensor and axial-vector resonant contributions, while other processes receive only    axial-vector resonant contributions.  In cases where both kinds of resonant contributions exist, the axial-vector contributions are dominant. Some branching ratios with  axial-vector resonance states are large, and they  may be measured experimentally in the near future. In addition, the sensitivities of the branching ratios of $D \to A \ell^+ \nu_\ell $ and $A \to VP$ decays  to the parameters are also investigated, and   some decay branching ratios are found to be sensitive to the non-perturbative parameters.

\end{abstract}
\maketitle

\section{INTRODUCTION}

Semileptonic decays of the charmed mesons are very useful for determining the quark mixing parameters and values of the Cabibbo-Kobayashi-Maskawa (CKM) matrix elements, and in searching for new physics beyond the standard model.
The intermediate resonances in these semileptonic decays provide  a clean environment in which to explore meson spectroscopy, as there is  no interference from other particles \cite{Ke:2023qzc}.
Among  the semileptonic four-body decays $D \to VP\ell^+\nu_\ell$,   only  the branching ratios of  $D^+\to b^0_1(b_1^0\to \omega\pi^0)e^+\nu_e$ and $D^0\to b^-_1(b_1^-\to \omega\pi^-)e^+\nu_e$ decays  have experimental  upper limits \cite{PDG2022}.  Nevertheless, two semileptonic three-body decays $D^0\to K_1(1270)^-e^+\nu_e$ and $D^+\to\overline{K}_1(1270)^0e^+\nu_e$ have been found  by the CLEO Collaboration \cite{CLEO:2007oer}, and they have been much improved by  the BESIII Collaboration \cite{BESIII:2019eao,BESIII:2021uqr}.  The six non-leptonic decays $K_1 \to VP$  were measured by the ACCMOR Collaboration \cite{ACCMOR:1981yww}.  One can use them  to study the $D \to VP\ell^+\nu_\ell$ semileptonic decays with the axial-vector resonance states. As for the  $D \to VP\ell^+\nu_\ell$ decays with the tensor resonance states, only four $T \to VP$ modes have been measured \cite{PDG2022}; as there are no experimental data in either $D\to T\ell^+\nu_\ell$ or $D \to T(T \to VP)\ell^+\nu_\ell$ decays,  they can be studied by combining  theoretical calculations.

For semileptonic decays, since the leptons do not participate in the strong interaction, the weak and strong dynamics can be separated, and 
the theoretical description of the semileptonic decays is relatively simple.  The hadronic
transition form factors contain all the strong dynamics in the initial and final hadrons, and they are crucial for testing the theoretical calculations of the involved strong interaction.
The structure of the $D\to VP$ form factors
is more complicated than for the $D\to V$ or $D\to P$ form factors; nevertheless, we have not found their systematic calculations up to  now.
In the absence of reliable calculations,  symmetry analysis can provide very useful information about the decays.
Because SU(3) flavor symmetry  is independent of the detailed dynamics, it can provide us an opportunity to relate different
decay modes. Even though SU(3) flavor symmetry is only an approximate symmetry, it still provides  some valuable information about
the decays. SU(3) flavor symmetry has been widely used to study hadron decays, including b-hadron decays \cite{He:1998rq,He:2000ys,Fu:2003fy,Hsiao:2015iiu,He:2015fwa,He:2015fsa,Deshpande:1994ii,Gronau:1994rj,Gronau:1995hm,Shivashankara:2015cta,Zhou:2016jkv,Cheng:2014rfa,Wang:2021uzi,Wang:2020wxn} and
c-hadron decays \cite{Wang:2021uzi,Wang:2020wxn,Grossman:2012ry,Pirtskhalava:2011va,Cheng:2012xb,Savage:1989qr,Savage:1991wu,Altarelli:1975ye,Lu:2016ogy,Geng:2017esc,Geng:2018plk,Geng:2017mxn,Geng:2019bfz,Wang:2017azm,Wang:2019dls,Wang:2017gxe,Muller:2015lua}.

 Semileptonic  $D\to M\ell^+\nu_\ell$ decays have been studied, for example,  by the  quantum chromodynamics (QCD) sum rules approach \cite{Momeni:2020zrb,Momeni:2019uag,Zuo:2016msr,Hu:2021lkl,Khosravi:2008jw},   light-cone sum rules \cite{Huang:2021owr},   and  quark models \cite{Isgur:1988gb,Scora:1995ty,Cheng:2017pcq}.  The non-leptonic $M\to VP$  decays  have been studied by many methods, for examples,  the chiral unitary approach \cite{Geng:2006yb,Cabrera:2009ep}, quark models \cite{Ackleh:1996yt,Bloch:1999vka,Du:2021zdg}, and the effective Lagrangian of the strong interaction \cite{Lichard:2010zza,Li:1995aw,Li:1995tv,Wess:1967jq,Roca:2003uk,Vojik:2010ua}.
 The  four-body semileptonic  $D \to A(A \to VP)\ell^+\nu_\ell$ decays  have also been studied by the unitary extensions of chiral perturbation theory \cite{Wang:2020pyy}.
In this work, we will  analyze the $D \to M(M \to VP)\ell^+\nu_\ell$ four-body semileptonic decays  using SU(3) flavor symmetry/breaking approach.
We will first obtain the decay amplitude relations  of the $D \to M \ell^+ \nu_\ell$ or the vertex coefficient relations of $M \to VP$ by using  SU(3) flavor symmetry/breaking.
Then, after ignoring the SU(3) flavor  breaking effects due to poor experimental data of $D\to M\ell^+\nu_\ell$, $M\to VP$  and $D \to M(M \to VP)\ell^+\nu_\ell$,    the relevant non-perturbative coefficients and  mixing angles will be constrained   by the present experimental data. Finally,   the branching ratios which have not yet been measured will be predicted by the SU(3) flavor symmetry.  In addition,  the sensitivity of $\mathcal{B}(D \to A \ell^+ \nu_\ell )$ and $\mathcal{B}(A \to VP)$ decays  to the non-perturbative parameters will be displayed.

This paper is organized as follows. In Sect. \ref{Sec:D2PVlvA}, the $D\rightarrow A\ell^+\nu_\ell $, $A\to VP$ and $D\rightarrow A(A\to PV)\ell^+\nu_\ell $ decays are presented.  In Sect. \ref{Sec:D2PVlvT}, the $D\rightarrow T\ell^+\nu_\ell $, $T\to VP$ and $D\rightarrow T(T\to PV)\ell^+\nu_\ell $ decays are discussed.  We summarize our results in Sect. \ref{sec:Summary}.
Finally,  the meson multiplets of the SU(3) flavor group   are listed in the  Appendix.

\section{Semileptonic $D\rightarrow A(A\to PV)\ell^+\nu_\ell $ decays} \label{Sec:D2PVlvA}
\subsection{Semileptonic $D\rightarrow A\ell^+\nu_\ell $ decays} \label{Sec:D2Alv}

\subsubsection{Theoretical framework for the $D\to A \ell^+\nu_\ell$ decays}

For the semileptonic decays of the $c\rightarrow q\ell^+\nu_\ell$ transition, the effective Hamiltonian can be expressed as
\begin{eqnarray}
\mathcal{H}_{eff}(c\rightarrow q\ell^+\nu_\ell)&=&\frac{G_F}{\sqrt{2}}V_{cq}^*\bar{q}\gamma^\mu(1-\gamma_5)c~\bar{\nu}_\ell\gamma_\mu(1-\gamma_5)\ell,\label{Heff}
\end{eqnarray}
where $G_F$ is the Fermi constant, $V_{cq}$ is the CKM matrix element with $q=d,s$.
The decay amplitudes of the $D\to A\ell^+\nu_\ell$ decays can be written as \cite{Wang:2022yyn}
\begin{eqnarray}
\mathcal{M}(D\rightarrow A\ell^+\nu_\ell)&=&\frac{G_F}{\sqrt{2}}\sum_{m,n}g_{mn} L^{\lambda_\ell\lambda_\nu}_mH^{\lambda_A}_{n}.\label{Eq:MD2Alv1}
\end{eqnarray}
Leptonic amplitudes $L^{\lambda_\ell\lambda_\nu}_m$ and hadronic amplitudes $H^{\lambda_{A}}_{n}$  are represented as
\begin{eqnarray}
L^{\lambda_\ell\lambda_\nu}_m&=&\epsilon_{\alpha}(m)\bar{\nu_\ell}\gamma^{\alpha}(1-\gamma_5)\ell, \label{Eq:MD2Alv2}\\
H^{\lambda_{A}}_{n}&=&
V_{cq}^*\epsilon_{\beta}^*(n)\langle{A}(p,\varepsilon^*)|\bar{q}\gamma^\beta(1-\gamma_5)c|D(p_D)\rangle, \label{Eq:MD2Alv3}
\end{eqnarray}
where $\lambda_A=0,\pm1$, $\lambda_\ell=\pm\frac{1}{2}$, and $\lambda_\nu=+\frac{1}{2}$ are the particle helicity, and $\epsilon(m)$  with $m=0,t,\pm1$ and $\varepsilon^*$  are the polarization  vectors of the virtual $W$ and axial-vector mesons $A$, respectively.

The hadron amplitudes can be parameterized by  form factors of the $D\to A$ transition  \cite{Cheng:2003sm}
\begin{eqnarray}
\left<A(p,\varepsilon^*)\left|\bar{q}\gamma_{\mu}(1-\gamma_5)c\right|D(p_D)\right>
&=&\frac{2iA(q^2)}{m_D-m_A}\epsilon_{\mu\nu\alpha\beta}\varepsilon^{*\nu}p^\alpha_Dp^\beta\nonumber\\
&&-\left[\varepsilon^*_\mu(m_D-m_A)V_1(q^2)-(p_D+p)_\mu(\varepsilon^*.p_D)\frac{V_2(q^2)}{m_D-m_A}\right]\nonumber\\
&&+q_\mu(\varepsilon^*.p_D)\frac{2m_A}{q^2}[V_3(q^2)-V_0(q^2)],
\end{eqnarray}
where $q^2\equiv(p_D-p)^2$, $A(q^2)$ and $V_{0,1,2,3}(q^2)$ are the form factors of the $D\to A$ transition,  and $m_{A,D,\ell}$ are the masses of $A,D,\ell$.
Ten hadronic amplitudes can then be written as \cite{Sun:2011ssd,Li:2009tx,Colangelo:2019axi}
\begin{eqnarray}
H_{\pm} &=&(m_{D}-m_A)V_1(q^2)\mp\frac{2m_{D}|\vec{p}_A|}{(m_{D}-m_A)}A(q^2), \\
H_{0}&=& \frac{1}{2m_A\sqrt{q^2}}\left[(m_{D}^2-m_A^2-q^2)(m_{D}-m_A)V_1(q^2)-\frac{4m_{D}^2|\vec{p}_A|^2}{m_{D}-m_A}V_2(q^2)\right], \\
H_{t}&=& \frac{2m_{D}|\vec{p}_A|}{\sqrt{q^2}}V_0(q^2),
\end{eqnarray}
in which  $|\vec{p}_A|\equiv\sqrt{\lambda(m_{D}^2,m_A^2,q^2)}/(2m_{D})$ with $\lambda(a,b,c)=a^2+b^2+c^2-2ab-2ac-2bc$.

The differential branching ratios are  \cite{Ivanov:2019nqd}
\begin{eqnarray}
&& \frac{d\mathcal{B}(D\to A\ell^+\nu_\ell)}{dq^2}=\frac{\tau_{D}G_F^2 |V_{cq}|^2\lambda^{1/2}(q^2-m_\ell^2)^2}{24(2\pi)^3m_{D}^3q^2} {\cal H}_{\rm total}, \label{eq:dbr}\\
\mbox{with}~~ &&{\cal H}_{\rm total}\equiv ({\cal H}_U+{\cal   H}_L)\left(1+\frac{m_\ell^2}{2q^2}\right) +\frac{3m_\ell^2}{2q^2}{\cal H}_S, \label{eq:htotal}\\
\mbox{and}~~&&{\cal H}_U=|H_+|^2+|H_-|^2, \quad {\cal H}_L=|H_0|^2, \quad {\cal   H}_P=|H_+|^2-|H_-|^2,\quad    {\cal H}_S=|H_t|^2, \quad {\cal H}_{SL}=\Re(H_0H_t^{\dag}),  \label{eq:hh}
\end{eqnarray}
where  $\lambda\equiv
\lambda(m_{D}^2,m_A^2,q^2)$,  $m_\ell^2\leq q^2\leq(m_{D}-m_A)^2$, and $\tau_{D}$ are the mean lifetime of $D$ meson.
The branching ratios are usually obtained by  Eq. (\ref{eq:dbr}) and the form factors; however, the results of the form factors are dependent on different approaches.

Branching ratios also can be obtained by using the available data and  the SU(3) flavor symmetry/breaking, which is independent of
the detailed dynamics, offering us an opportunity to relate different decay modes.   In the $\ell=e,\mu$ cases,  $m_\ell^2$ is small, and  it  can be ignored in ${\cal H}_{\rm total}$ in Eq. (\ref{eq:htotal}).
Then ${\cal H}_{\rm total}$ only  contains the hadronic information, which can be related by the  the SU(3) flavor symmetry/breaking.

The SU(3) flavor symmetry analysis can provide the hadronic helicity amplitude relationships among various decay models.
The representations for meson multiplets of the SU(3) flavor group are  collected in the Appendix.
Similar to the SU(3) flavor symmetry/breaking  hadronic helicity amplitudes of $D\to P/V/S\ell^+\nu_\ell$  decays given in Ref. \cite{Wang:2022yyn}, the hadronic helicity amplitudes of the $D\to A\ell^+\nu_\ell$  decays can be parameterized as
\begin{eqnarray}
H(D\to M \ell^+\nu_\ell) =c_{0}^{M} D_iM^i_jH^j+c_{1}^M D_kW_{i}^kM^i_jH^j +c_{2}^M D_iM_{k}^iW_{j}^kH^j, \label{Eq:AmpD2AlvSU3}
\end{eqnarray}
where $H^2\equiv V^{*}_{cd}$ and $H^3\equiv V^{*}_{cs}$  are the CKM matrix elements, $W$ is the SU(3) flavor breaking related matrix \cite{Xu:2013dta,He:2014xha},    $c_{0,1,2}^A$ and $c_{0,1,2}^B$  are the non-perturbative coefficients corresponding to $M=A$ and $M=B$, respectively,  $c_{0}^{A,B}$ are those   under the SU(3) flavor symmetry, and $c_{1,2}^{A,B}$  are the SU(3) flavor breaking ones. Note that the Okubo-Zweig-Iizuka suppressed processes, such as $ D^+_s\to b_1(1235)^0\ell^+\nu_\ell$ decays studied in Ref. \cite{BESIII:2023clm}, are not considered in this work.
In terms of Eq. (\ref{Eq:AmpD2AlvSU3}), the hadronic helicity amplitude relations for the $D\to A \ell^+\nu_\ell$ decays are summarized in Tab. \ref{Tab:HD2AlvAmp}.

From Tab. \ref{Tab:HD2AlvAmp},  one  can easily obtain the amplitude relations between different decay processes.   The CKM matrix elements $V_{cs}$ and $V_{cd}$  are also listed in Tab. \ref{Tab:HD2AlvAmp} for  convenient comparison. Since the light axial-vector matrix is distinguished as $A$ and $B$ in the meson multiplets, there are eight non-perturbative parameters  $A_{1}$, $A_{2}$, $A_{3}$, $A_{4}$ and $B_{1}$, $B_{2}$, $B_{3}$, $B_{4}$ as defined in the caption of Tab. \ref{Tab:HD2AlvAmp}. If we neglect  the SU(3) flavor breaking effects of $c^A_1$, $c^A_2$, $c^B_1$,  and $c^B_2$, there are only  two non-perturbative parameters  $c^A_0$ and $c^B_0$. The hadronic  amplitudes of various decay processes are connected to the non-perturbative parameters $c^A_0$ and $c^B_0$, as well as the mixing angles $\theta_{K_1}$, $\theta_{1P_1}$ and $\theta_{3P_1}$.

\begin{table}[thb]
\renewcommand\arraystretch{1.3}
\tabcolsep 0.1in
\centering
\caption{The hadronic amplitudes of the $D\to A\ell^+\nu_\ell$ decays. $A_1\equiv c^A_0+c^A_1-2c^A_2$, $A_2\equiv c^A_0-2c^A_1-2c^A_2$, $A_3\equiv c^A_0+c^A_1+c^A_2$, $A_4\equiv c^A_0-2c^A_1+c^A_2$. $B_1\equiv c^B_0+c^B_1-2c^B_2$, $B_2\equiv c^B_0-2c^B_1-2c^B_2$, $B_3\equiv c^B_0+c^B_1+c^B_2$, $B_4\equiv c^B_0-2c^B_1+c^B_2$. $A_1=A_2=A_3=A_4=c^A_0$ and $B_1=B_2=B_3=B_4=c^B_0$ if  neglecting the SU(3) flavor breaking $c^A_1$, $c^A_2$, and $c^B_1$, $c^B_2$ terms. }\vspace{0.1cm}
{\footnotesize
\begin{tabular}{lc|lc}  \hline\hline
 ~~~~Decay modes & SU(3) hadronic amplitudes & ~~~~Decay modes & SU(3) hadronic amplitudes\\\hline
$D^0\to K_1(1270)^-\ell^+\nu_\ell$&$(sin\theta_{K_1}A_1+cos\theta_{K_1}B_1)~V^*_{cs}$                                         &$D^0\to K_1(1400)^-\ell^+\nu_\ell$&$(cos\theta_{K_1}A_1-sin\theta_{K_1}B_1)~V^*_{cs}$\\\hline
$D^+\to \overline{K}_1(1270)^0\ell^+\nu_\ell$&$(sin\theta_{K_1}A_1+cos\theta_{K_1}B_1)~V^*_{cs}$                              &$D^+\to \overline{K}_1(1400)^0\ell^+\nu_\ell$&$(cos\theta_{K_1}A_1-sin\theta_{K_1}B_1)~V^*_{cs}$\\\hline
$D^+_s\to f_1(1285)\ell^+\nu_\ell$&$(\frac{1}{\sqrt{3}}cos\theta_{3P_1}-\sqrt{\frac{2}{3}}sin\theta_{3P_1})~A_2~V^*_{cs}$     &$D^+_s\to f_1(1420)\ell^+\nu_\ell$&$(-\frac{1}{\sqrt{3}}sin\theta_{3P_1}-\sqrt{\frac{2}{3}}cos\theta_{3P_1})~A_2~V^*_{cs}$\\\hline
$D^+_s\to h_1(1170)\ell^+\nu_\ell$&$(\frac{1}{\sqrt{3}}cos\theta_{1P_1}-\sqrt{\frac{2}{3}}sin\theta_{1P_1})~B_2~V^*_{cs}$     &$D^+_s\to h_1(1415)\ell^+\nu_\ell$&$(-\frac{1}{\sqrt{3}}sin\theta_{1P_1}-\sqrt{\frac{2}{3}}cos\theta_{1P_1})~B_2~V^*_{cs}$\\\hline
$D^0\to a_1(1260)^-\ell^+\nu_\ell$&$A_3~V^*_{cd}$                                                                             &$D^0\to b_1(1235)^-\ell^+\nu_\ell$&$B_3~V^*_{cd}$\\\hline
$D^+\to a_1(1260)^0\ell^+\nu_\ell$&$-\frac{1}{\sqrt{2}}~A_3~V^*_{cd}$                                                          &$D^+\to b_1(1235)^0\ell^+\nu_\ell$&$-\frac{1}{\sqrt{2}}~B_3~V^*_{cd}$\\\hline
$D^+\to f_1(1285)\ell^+\nu_\ell$&$(\frac{1}{\sqrt{3}}cos\theta_{3P_1}+\frac{1}{\sqrt{6}}sin\theta_{3P_1})~A_3~V^*_{cd}$         &$D^+\to f_1(1420)\ell^+\nu_\ell$&$(-\frac{1}{\sqrt{3}}sin\theta_{3P_1}+\frac{1}{\sqrt{6}}cos\theta_{3P_1})~A_3~V^*_{cd}$\\\hline
$D^+\to h_1(1170)\ell^+\nu_\ell$&$(\frac{1}{\sqrt{3}}cos\theta_{1P_1}+\frac{1}{\sqrt{6}}sin\theta_{1P_1})~B_3~V^*_{cd}$         &$D^+\to h_1(1415)\ell^+\nu_\ell$&$(-\frac{1}{\sqrt{3}}sin\theta_{1P_1}+\frac{1}{\sqrt{6}}cos\theta_{1P_1})~B_3~V^*_{cd}$\\\hline
$D^+_s\to K_1(1270)^0\ell^+\nu_\ell)$&$(sin\theta_{K_1}A_4+cos\theta_{K_1}B_4)~V^*_{cd}$                                        &$D^+_s\to K_1(1400)^0\ell^+\nu_\ell)$&$(cos\theta_{K_1}A_4-sin\theta_{K_1}B_4)~V^*_{cd}$\\\hline
\end{tabular}}\label{Tab:HD2AlvAmp}
\end{table}


\subsubsection{Numerical results for the $D\to A \ell^+\nu_\ell$ decays}
In our numerical analysis, the theoretical input parameters and experimental data  within $1\sigma$ error and  within  $2\sigma$ errors  are taken from the PDG \cite{PDG2022}.  In the $D\to A \ell^+\nu_\ell$ decays, only $D^0\to K_1(1270)^-e^+\nu_e$ and $D^+\to\overline{K}_1(1270)^0e^+\nu_e$ decays
have been measured, and the branching ratio of  $D^+_s\to \overline{K}_1(1270)^0e^+\nu_e$ is constrained by an  upper limit.  Their experimental data with $1\sigma$ error  are \cite{PDG2022,BESIII:2023clm}
\begin{eqnarray}
\mathcal{B}(D^0\to K_1(1270)^-e^+\nu_e)&=&(1.01\pm0.18)\times10^{-3},\label{Eq:D2K1lvdata1}\\
\mathcal{B}(D^+\to \overline{K}_1(1270)^0e^+\nu_e)&=&(2.30^{+0.40}_{-0.42})\times10^{-3},\label{Eq:D2K1lvdata1p}\\
\mathcal{B}(D^+_s\to \overline{K}_1(1270)^0e^+\nu_e)&<&4.1\times10^{-4} ~~\mbox{at 90\% CL}.\label{Eq:D2K1lvdata2}
\end{eqnarray}
 Only two experimental data can not constrain the five parameters $c^A_0$ and $c^B_0$, $\theta_{K_1}$, $\theta_{1P_1}$, and $\theta_{3P_1}$.  Three mixing angles $\theta_{K_1}$, $\theta_{1P_1}$, and $\theta_{3_{P_1}}$ can also be constrained by the $A\to PV$ non-leptonic decays, which will be discussed in the next section.
$\mathcal{B}(D\to a^-_1/a^0_1\ell^+\nu_\ell)$ and $\mathcal{B}(D\to b^-_1/b^0_1\ell^+\nu_\ell)$ are proportional to $|c^A_0|^2$ and $|c^B_0|^2$, respectively.
As given in Refs. \cite{Momeni:2019uag,Momeni:2022gqb},  the theoretical predictions of $\mathcal{B}(D\to a^-_1/a^0_1\ell^+\nu_l)$ and $\mathcal{B}(D\to b^-_1/b^0_1\ell^+\nu_l)$ are quite dependent on different  approaches. For the sake of conservatism, we only use $\Big|\frac{c^A_0}{c^B_0}\Big|=\sqrt{\frac{\mathcal{B}(D\to a^-_1/a^0_1\ell^+\nu_l)}{\mathcal{B}(D\to b^-_1/b^0_1\ell^+\nu_l)}}$ from Refs. \cite{Momeni:2019uag,Momeni:2022gqb}, which lies within $[0.88,1.97]$.

Now, using the experimental data for $\mathcal{B}(D^0\to K_1(1270)^-e^+\nu_e)$, $\mathcal{B}(D^+\to\overline{K}_1(1270)^0e^+\nu_e)$, and $\mathcal{B}(D^+_s\to \overline{K}_1(1270)^0e^+\nu_e)$  in Eqs. (\ref{Eq:D2K1lvdata1}-\ref{Eq:D2K1lvdata2}), the ratio $\Big|\frac{c^A_0}{c^B_0}\Big|\in[0.88,1.97]$ from Refs. \cite{Momeni:2019uag,Momeni:2022gqb}, and  the relevant constraints from later $\mathcal{B}(A\to PV)$ in Eqs. (\ref{Eq:K12702VPdata}-\ref{Eq:K14002VPdata}) and $\mathcal{B}(D\to Ae^+\nu_e,A\to PV)$ in Eqs. (\ref{Eq:BrB2PVlvA1}-\ref{Eq:BrB2PVlvA2}), in terms of the hadronic amplitude relations listed in Tab. \ref{Tab:HD2AlvAmp},  one can constrain relevant parameters and then provide the predictions for $\mathcal{B}(D\to A\ell^+\nu_\ell)$,  which have not yet been measured.

\begin{figure}[t]
\begin{center}
\includegraphics[scale=0.8]{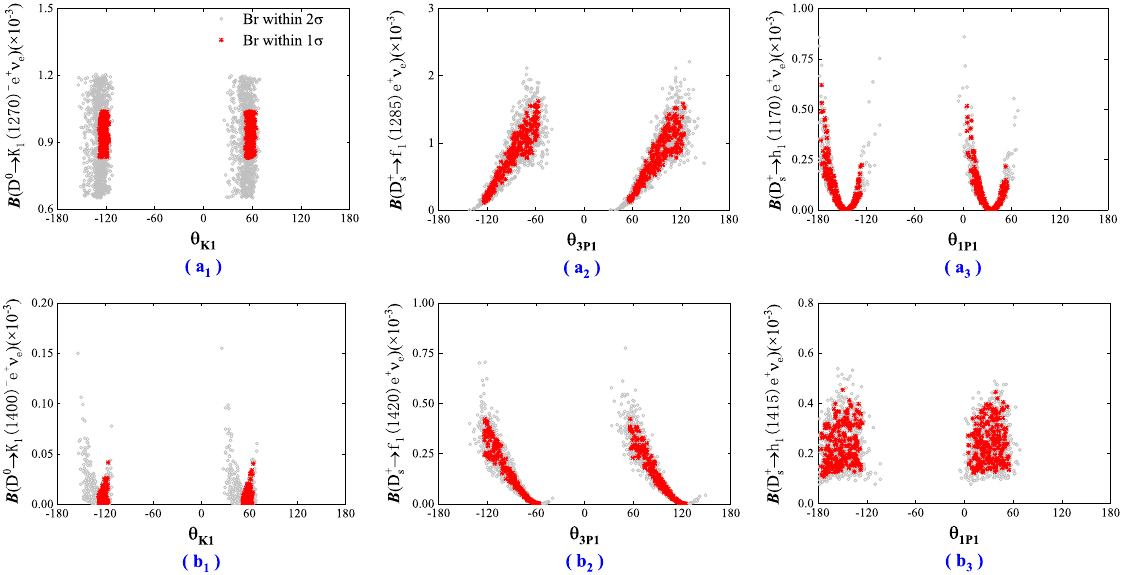}
\end{center}
\caption{The allowed ranges of  three mixing angles $\theta_{K_1}$, $\theta_{3P_1}$, and $\theta_{1P_1}$ from $\mathcal{B}(B\to Ae^+\nu_e)$, $\mathcal{B}(A\to PV)$, and $\mathcal{B}(D\to Ae^+\nu_e,A\to PV)$. The gray regions show the results with the theoretical input parameters and the experimental data within 2$\sigma$ errors, the red regions show  them within 1$\sigma$ error.}\label{fig:tworanges}
\end{figure}

We find that the  range of  $\Big|\frac{c^A_0}{c^B_0}\Big|$ is not reduced by the present experimental data.
As for  the three mixing angles, their constraints mainly come from the $A\to PV$ decays, and  they  are all constrained from $[-180^\circ,180^\circ]$ to two allowed ranges  as shown in Fig. \ref{fig:tworanges}.
An angle in one allowed range can be found as a supplementary angle in another allowed range, and the branching ratios have similar sensitivity to the angles of the two ranges.
Therefore,  we only discuss one of two ranges,  which are about larger than $0^\circ$, and the allowed ranges are
\begin{eqnarray}
&c^A_0\in[1.44, 3.00],~c^B_0\in[0.98, 2.27],~\theta_{K_1}\in[26^\circ, 70^\circ], ~\theta_{1P_1}\in[-15^\circ, 75^\circ],~ \theta_{3P_1}\in[32^\circ, 149^\circ] ~\mbox{within}~2\sigma ~\mbox{errors},\label{Eq:theta1}\\
&c^A_0\in[1.69, 2.42],~c^B_0\in[1.12, 2.08],~\theta_{K_1}\in[52^\circ, 65^\circ], ~\theta_{1P_1}\in[5^\circ, 56^\circ],~ \theta_{3P_1}\in[56^\circ, 125^\circ]~\mbox{within}~ 1\sigma ~\mbox{error}.~~~~~~\label{Eq:theta2}
\end{eqnarray}
As previously mentioned, there are several earlier estimations on the mixing angle $\theta_{K_1}$.   The allowed $\theta_{K_1}$ is $[35^\circ,55^\circ]$ in terms of two parameters, the mass difference of the $a_1(1260)$ and
$b_1(1235)$ mesons and the ratio of the constituent quark masses \cite{Burakovsky:1997dd};  $33^\circ$ and $57^\circ$ by $\tau$ decay \cite{Suzuki:1993yc},  $(34\pm13)^\circ$ by $B\to K_1\gamma$ \cite{Hatanaka:2008xj}; $33^\circ$ is much more favored than $57^\circ$ in Ref. \cite{Cheng:2013cwa}, etc.
The allowed ranges of $\theta_{1P_1}$ and $\theta_{3P_1}$ are obtained, $\theta_{1P_1}=(23.3\pm1)^\circ$ or $(38.3\pm1)^\circ$ \cite{Cheng:2011pb},   $\theta_{3P_1}=(4.3\pm2)^\circ$ or $(66.3\pm2)^\circ$ \cite{Cheng:2011pb},  $\theta_{3_{P_1}}=\big(19.4^{+4.5}_{-4.6}\big)^\circ$ or $\big(51.1^{+4.5}_{-4.6}\big)^\circ$  \cite{Yang:2010ah}.  Our bounds  within $1\sigma$ error bar favor the large values of   $\theta_{K_1}$ and $\theta_{3_{P_1}}$ in Refs. \cite{Cheng:2011pb,Yang:2010ah}.

\begin{table}[t]
\renewcommand\arraystretch{1.5}
\tabcolsep 0.1in
\centering
\caption{ The branching ratio predictions of the $D\to A\ell^+\nu_\ell$ decays within $1\sigma $ error, and $^E$denotes experimental data.  }\vspace{0.1cm}
{\scriptsize
\begin{tabular}{lcl|cl}  \hline\hline
                                                                          &Our results $(\ell=e)$       & Previous results  $(\ell=e)$                          &Our results $(\ell=\mu)$       & Previous results  $(\ell=\mu)$                                                                       \\ \hline
$\mathcal{B}(D^0\to K_1(1270)^-\ell^+\nu_\ell)(\times10^{-3})$                    &$^{0.93\pm0.10}_{1.01\pm0.18^E}$                    &$ \cdots$                                                                   &$0.81\pm0.09$                    &$\cdots  $                                                                                        \\
$\mathcal{B}(D^0\to K_1(1400)^-\ell^+\nu_\ell)(\times10^{-3})$                    &$0.02\pm0.02$                    &$ \cdots$                                                                   &$0.02\pm0.02$                    &$\cdots $                                                                                         \\ \hline
$\mathcal{B}(D^+\to \overline{K}_1(1270)^0\ell^+\nu_\ell)(\times10^{-3})$         &$^{2.43\pm0.27}_{2.30^{+0.40~E}_{-0.42}}$                    &$3.2\pm0.40$ \cite{Cheng:2017pcq}                                           &$2.10\pm0.23$                    &$2.6\pm0.30$ \cite{Cheng:2017pcq}                                                           \\
$\mathcal{B}(D^+\to \overline{K}_1(1400)^0\ell^+\nu_\ell)(\times10^{-3})$         &$0.05\pm0.05$                    &$\{0.005,0.02\}$ \cite{Cheng:2017pcq}                                       &$0.04\pm0.04$                    &$\{0.004,0.017\}$ \cite{Cheng:2017pcq}                                                      \\ \hline
$\mathcal{B}(D^+_s\to f_1(1285)\ell^+\nu_\ell)(\times10^{-3})$                    &$0.86\pm0.73$                    &$\{0.06,0.36\}$ \cite{Cheng:2017pcq}                                        &$0.76\pm0.65$                    &$\{0.052,0.306\}$ \cite{Cheng:2017pcq}                                                      \\
$\mathcal{B}(D^+_s\to f_1(1420)\ell^+\nu_\ell)(\times10^{-3})$                    &$0.21\pm0.21$                    &$0.25\pm0.05$ \cite{Cheng:2017pcq}                                          &$0.18\pm0.18$                    &$0.21\pm0.05$ \cite{Cheng:2017pcq}                                                          \\ \hline
$\mathcal{B}(D^+_s\to h_1(1170)\ell^+\nu_\ell)(\times10^{-3})$                    &$0.26\pm0.26$                    &$\{0,0.197\}$ \cite{Cheng:2017pcq}                                          &$0.24\pm0.24$                    &$\{0,0.174\}$ \cite{Cheng:2017pcq}                                                           \\
$\mathcal{B}(D^+_s\to h_1(1415)\ell^+\nu_\ell)(\times10^{-3})$                    &$0.28\pm0.16$                    &$0.64\pm0.07$ \cite{Cheng:2017pcq}                                          &$0.24\pm0.14$                    &$0.54\pm0.06$ \cite{Cheng:2017pcq}                                                            \\ \hline
$\mathcal{B}(D^0\to a_1(1260)^-\ell^+\nu_\ell)(\times10^{-5})$                    &$4.46\pm2.32$                    &$6.90$ \cite{Huang:2021owr}, $5.261^{+0.745}_{-0.639}$ \cite{Hu:2021lkl}    &$3.93\pm2.08$                    &$6.27$ \cite{Huang:2021owr} , $4.732^{+0.685}_{-0.590}$ \cite{Hu:2021lkl}                                                    \\
$\mathcal{B}(D^0\to b_1(1235)^-\ell^+\nu_\ell)(\times10^{-5})$                    &$2.61\pm1.44$                    &$4.85$ \cite{Huang:2021owr}                                                 &$2.29\pm1.26$                    &$4.40$ \cite{Huang:2021owr}                                                                     \\ \hline
$\mathcal{B}(D^+\to a_1(1260)^0\ell^+\nu_\ell)(\times10^{-5})$                    &$5.79\pm3.00$                    &$9.38$ \cite{Huang:2021owr}, $6.673^{+0.947}_{-0.811}$ \cite{Hu:2021lkl}    &$5.11\pm2.70$                    &$8.52$ \cite{Huang:2021owr} , $6.002^{+0.796}_{-0.748}$ \cite{Hu:2021lkl}                                                      \\
$\mathcal{B}(D^+\to b_1(1235)^0\ell^+\nu_\ell)(\times10^{-5})$                    &$3.41\pm1.88$                    &$7.4\pm0.70$ \cite{Cheng:2017pcq}, $6.58$ \cite{Huang:2021owr}              &$2.99\pm1.65$                    &$6.4\pm0.6$ \cite{Cheng:2017pcq}, $6.00$ \cite{Huang:2021owr}                                                      \\ \hline
$\mathcal{B}(D^+\to f_1(1285)\ell^+\nu_\ell)(\times10^{-5})$                      &$1.88\pm1.88$                    &$3.7\pm0.80$ \cite{Cheng:2017pcq}                                           &$1.61\pm1.61$                    &$3.2\pm0.6$ \cite{Cheng:2017pcq}                                                                 \\
$\mathcal{B}(D^+\to f_1(1420)\ell^+\nu_\ell)(\times10^{-5})$                      &$0.64\pm0.54$                    &$\{0.02,0.14\}$ \cite{Cheng:2017pcq}                                        &$0.48\pm0.41$                    &$\{0.02,0.12\}$ \cite{Cheng:2017pcq}                                                               \\ \hline
$\mathcal{B}(D^+\to h_1(1170)\ell^+\nu_\ell)(\times10^{-5})$                      &$5.28\pm3.00$                    &$14\pm1.50$ \cite{Cheng:2017pcq}                                            &$4.73\pm2.69$                    &$12.2\pm1.3$ \cite{Cheng:2017pcq}                                                                  \\
$\mathcal{B}(D^+\to h_1(1415)\ell^+\nu_\ell)(\times10^{-5})$                      &$0.11\pm0.11$                    &$\{0,0.02\}$ \cite{Cheng:2017pcq}                                           &$0.08\pm0.08$                    &$\{0,0.02\}$ \cite{Cheng:2017pcq}                                                                 \\ \hline
$\mathcal{B}(D^+_s\to K_1(1270)^0\ell^+\nu_\ell)(\times10^{-5})$                  &$^{11.77\pm1.39}_{<41^E}$                   &$17\pm2.00$ \cite{Cheng:2017pcq}                                            &$10.57\pm1.25$                   &$15\pm2$ \cite{Cheng:2017pcq}                                                                      \\
$\mathcal{B}(D^+_s\to K_1(1400)^0\ell^+\nu_\ell)(\times10^{-5})$                  &$0.32\pm0.32$                    &$\{0.05,0.14\}$ \cite{Cheng:2017pcq}                                        &$0.27\pm0.27$                    &$\{0.05,0.12\}$ \cite{Cheng:2017pcq}                                                               \\ \hline
\end{tabular}}\label{Tab:BrD2Alv}
\end{table}

Our numerical predictions  within 1$\sigma$ error for  the branching ratios of the $D\to A\ell^+\nu_\ell$ decays  are listed in Tab. \ref{Tab:BrD2Alv}, and  the experimental data given in Eqs. (\ref{Eq:D2K1lvdata1}-\ref{Eq:D2K1lvdata2}) are also listed for  the convenience of comparison.
$\mathcal{B}(D\to K_1\ell^+\nu_\ell)$ and $\mathcal{B}(D_s\to f_1/h_1\ell^+\nu_\ell)$  are  proportional to $|V^*_{cs}|^2\approx0.95$.  $\mathcal{B}(D\to f_1/h_1/a_1/b_1\ell^+\nu_\ell)$ and $\mathcal{B}(D_s\to K_1\ell^+\nu_\ell)$ are  proportional to $|V^*_{cd}|^2\approx0.05$.  As given in Tab. \ref{Tab:BrD2Alv}, $\mathcal{B}(D\to K_1(1270)\ell^+\nu_\ell)$ and $\mathcal{B}(D_s\to f_1(1285)/f_1(1420)/h_1(1170)/h_1(14115)\ell^+\nu_\ell)$ are
 on the order of $\mathcal{O}(10^{-4}-10^{-3})$, which may be measured experimentally in the near future.    $\mathcal{B}(D\to K_1(1400)\ell^+\nu_\ell)$ are strongly suppressed by the mixing angle $\theta_{K_1}$. Other branching ratios are on the order of $\mathcal{O}(10^{-6}-10^{-5})$.
Previous predictions from Refs. \cite{Cheng:2017pcq,Huang:2021owr} are also listed in the last column of Tab. \ref{Tab:BrD2Alv}.
Most of our predictions are highly coincident  with them within 1$\sigma$ error, and some are coincident within 2$\sigma$ errors.
Nevertheless, our predictions of $\mathcal{B}(D^+\to b_1(1235)^0\ell^+\nu_\ell)$ and $\mathcal{B}(D^+\to h_1(1170)\ell^+\nu_\ell)$  are obviously smaller than those in Refs. \cite{Cheng:2017pcq,Huang:2021owr}.

\begin{figure}[b]
\begin{center}
\includegraphics[scale=0.55]{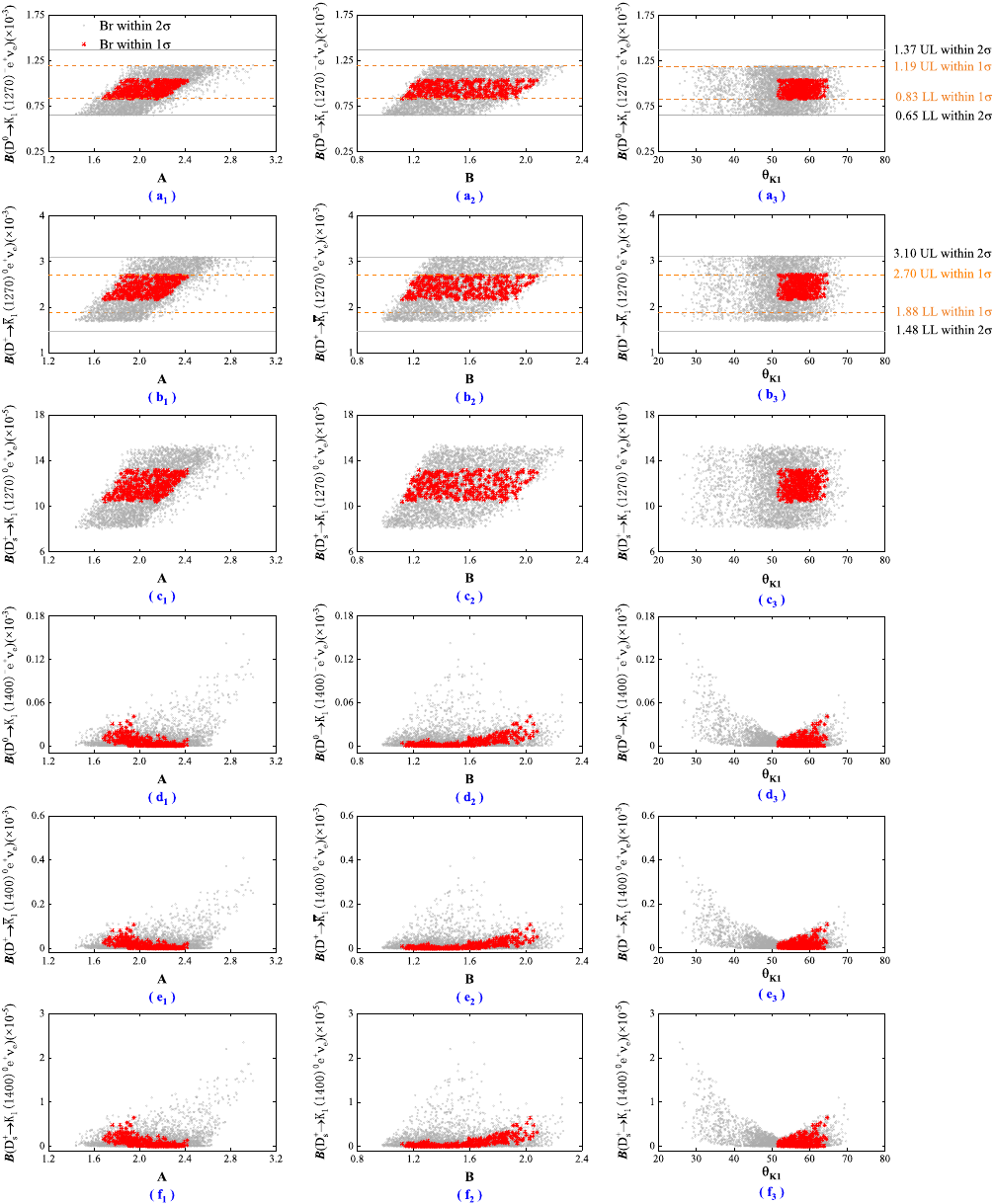}
\end{center}
\caption{Branching ratios of the $D\to K_1e^+\nu_e$ decays. The gray solid line indicates the experimental data within 2$\sigma$ errors,and  the red dotted line indicates the experimental data within 1$\sigma$ error. UL/LL stands for upper/lower limits of experimental data, the same in Figs. \ref{fig:A2VP1} and \ref{fig:A2VP2}.}\label{fig:BrD2AlvK1}
\begin{center}
\includegraphics[scale=0.55]{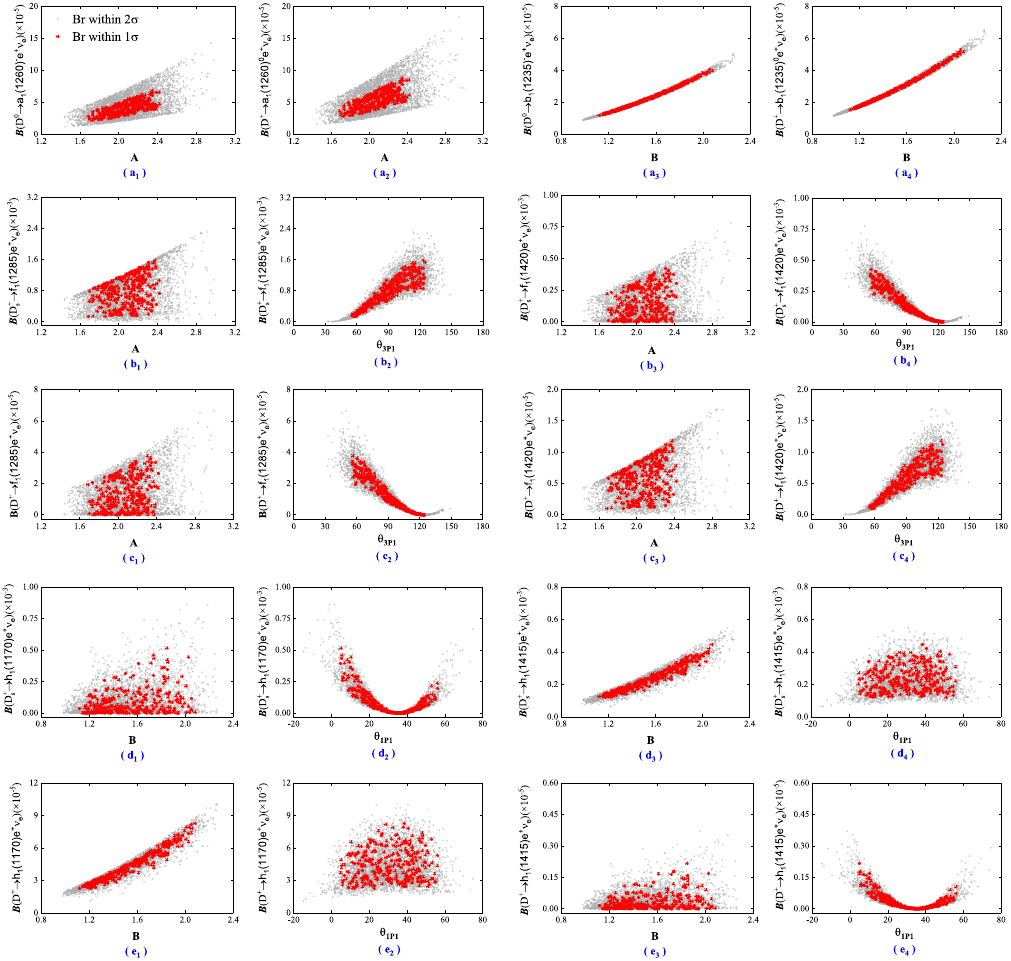}
\end{center}
\caption{Branching ratios of the $D_{(s)}\to a_1/b_1/f_1/h_1 e^+\nu_e$ decays. }\label{fig:BrD2Alvab1}
\end{figure}

In addition,  we show the sensitivity of the branching ratios to the non-perturbative   parameters $A$ and $B$ as well as  the mixing angles $\theta_{K_1}$, $\theta_{3P_1}$,  and $\theta_{1P_1}$ in Figs. \ref{fig:BrD2AlvK1} and \ref{fig:BrD2Alvab1}. We give the remarks as follows.
\begin{itemize}
\item As  given in Tab. \ref{Tab:HD2AlvAmp},  the hadronic amplitudes of  the $D^0 \to K_1(1270)^- e^+\nu_e$, $D^+\to \overline{K}_1(1270)^0e^+\nu_e$, and $D^+_s\to {K}_1(1270)^0e^+\nu_e$  decays  are  proportional to  $sin\theta_{K_1}c^A_0+cos\theta_{K_1}c^B_0$, and  the hadronic amplitudes of  the $D^0 \to K_1(1400)^- e^+\nu_e$, $D^+\to \overline{K}_1(1400)^0e^+\nu_e$, and $D^+_s\to {K}_1(1400)^0e^+\nu_e$  decays  are  proportional to  $cos\theta_{K_1}c^A_0-sin\theta_{K_1}c^B_0$.  Thus as shown in Fig. \ref{fig:BrD2AlvK1}, three $D \to K_1(1270) e^+\nu_e$ decays or three $D\to {K}_1(1400)e^+\nu_e$ decays  have similar variation trends as the three parameters  $c^A_0$, $c^B_0$, or $\theta_{K_1}$.  From Fig. \ref{fig:BrD2AlvK1} ($a_{1,2,3}$) and ($b_{1,2,3}$),  one  can see that the experimental  lower limit of $\mathcal{B}(D^0 \to K_1(1270)^- e^+\nu_e)$  and the experimental  upper limit of $\mathcal{B}(D^+\to \overline{K}_1(1270)^0e^+\nu_e)$ give  effective constraints on our predictions.  Three $\mathcal{B}(D \to K_1(1270) e^+\nu_e)$  in Fig. \ref{fig:BrD2AlvK1} ($a_{1,2,3}$), ($b_{1,2,3}$), and  ($c_{1,2,3}$) have some sensitivity to $A$ and $B$.   As shown in  Fig. \ref{fig:BrD2AlvK1} ($d_{1,2,3}$), ($e_{1,2,3}$), and  ($f_{1,2,3}$), three $\mathcal{B}(D\to {K}_1(1400)e^+\nu_e)$ are very sensitive to three parameters   $A$, $B$, and $\theta_{K_1}$.

\item As displayed in Fig. \ref{fig:BrD2Alvab1} ($a_{1-4}$), $\mathcal{B}(D \to a_1(1260) e^+\nu_e)$ and $D \to b_1(1235) e^+\nu_e)$ are only related to  $A$ and $B$, respectively,  and they are very sensitive to $A$ or $B$.
      Fig. \ref{fig:BrD2Alvab1} ($b_{1-4}$) and ($c_{1-4}$) show the $\mathcal{B}(D_{(s)} \to f_1(1285)/f_1(1420) e^+\nu_e)$ which are related to $A$ and $\theta_{3P_1}$.  Four  $\mathcal{B}(D_{(s)} \to f_1(1285)/f_1(1420) e^+\nu_e)$ have similar sensitivity to $A$, but the change trends of four branch ratios to $\theta_{3P_1}$ are different.
      Fig. \ref{fig:BrD2Alvab1} ($d_{1-4}$) and ($e_{1-4}$) show the $\mathcal{B}(D_{(s)} \to h_1(1170)/h_1(1415) e^+\nu_e)$, which are related to $B$ and $\theta_{1P_1}$.
      $\mathcal{B}(D^+ \to h_1(1170) e^+\nu_e)$ and $\mathcal{B}(D^+_s \to h_1(1415) e^+\nu_e)$ are sensitive to parameter $B$.   $\mathcal{B}(D^+_s \to h_1(1170) e^+\nu_e)$ and $\mathcal{B}(D^+ \to h_1(1415) e^+\nu_e)$ are sensitive to $\theta_{1P_1}$,  and  Fig. \ref{fig:BrD2Alvab1} ($d_{2}$) and ($e_{4}$) show that the branching ratios  first decrease  and then increase with  $\theta_{1P_1}$, reaching minimum values when $\theta_{1P_1}$ is about  $36^\circ$.

\end{itemize}


\subsection{Non-leptonic $A\to VP$ decays}\label{Sec:A2VP}
\subsubsection{Theoretical framework for the $A\to VP$ decays}

For the non-leptonic decays of axial-vector mesons  $A \to VP$, the decay branching ratios   can be written as \cite{Roca:2003uk,Parui:2022zfm}
\begin{equation}
\mathcal{B}(A\to VP)=\frac{\tau_A|\lambda_{AVP}|^2}{2\pi m_A^2}q'\left(1+\frac{2}{3}\frac{{q'}^2}{m_V^2}\right).
\label{eq:GAVP}
\end{equation}
where   $q'\equiv\frac{1}{2m_A}\lambda^{1/2}(m_A^2,m_V^2,m_P^2)$  is the momentum of the final state particles in the rest frame of the axial-vector meson, and  $\lambda_{AVP}$ are the coefficients of the $AVP$ vertex. $\lambda_{AVP}$  can be obtained from the effective Hamiltonian in Eq. (1) of Ref. \cite{Roca:2003uk} or the SU(3) flavor symmetry/breaking; we will use the latter to obtain $\lambda_{AVP}$  in this work.

There are small phase spaces for some axial-vector meson decays, such as $K_1(1270)\to \rho K/\omega K/\phi K/ K^*\eta$, $K_1(1400)\to \phi K/\phi\eta$, $a_1(1260)\to K^*K$, $b_1(1235)\to K^*K/\rho\eta$, $f_1(1285)\to K^*K$, $h_1(1170)\to K^*K/\omega \eta/\phi\eta$, and $h_1(1415)\to \phi \eta$.   The mass distributions of the resonances have been considered in Refs. \cite{Wang:2022vga,Roca:2021bxk}, and  the mass distributions of the resonances and vector mesons have been considered in Ref. \cite{Roca:2003uk}.  We will follow Ref. \cite{Roca:2003uk} to obtain our numerical results.
The decay branching ratios  with the mass distributions of the resonances and vector mesons  can be written as \cite{Roca:2003uk}
\begin{eqnarray} \nonumber
\mathcal{B}(A\to VP)&=&\frac{1}{N_AN_V\pi^2}\int ds_A d s_V Im \left\{\frac{1}{s_A-m_A^2+im_A\Gamma_A}\right\}
Im \left\{\frac{1}{s_V-m_V^2+im_V\Gamma_V}\right\}\\
&&\times \mathcal{B}_{AVP}(\sqrt{s_A},\sqrt{s_V})
\Theta(\sqrt{s_A}-\sqrt{s_V}-m_P)
\label{Eq:GammaA2VP2}
\end{eqnarray}
where $\Theta$ is the step function, $\mathcal{B}_{AVP}(\sqrt{s_A},\sqrt{s_V})=\frac{\tau_A|\lambda_{AVP}|^2}{2\pi s_A}q''\left(1+
\frac{2}{3}\frac{q''^2}{s_V}\right)$, with $q''\equiv\frac{1}{2\sqrt{s_A}}\lambda^{1/2}(s_A,s_V,m_P^2)$, and  $N_A$ and $N_V$ are  normalization factors used to account for some missing strength $ N_{A/V}\equiv (-\frac{1}{\pi})\int ds_{A/V} Im\big\{\frac{1}{s_{A/V}-m_{A/V}^2+im_{A/V}\Gamma_{A/V}}\big\}$  \cite{Roca:2021bxk}.  Following Ref. \cite{Wang:2022vga}, for  most small phase space decays, the upper limit $(m_{A/V}+n\Gamma_{A/V})^2$ and the lower limit $(m_{A/V}-n\Gamma_{A/V})^2$ with $n=2$ are used for $s_{A/V}$ in  Eq. (\ref{Eq:GammaA2VP2}). Nevertheless,  for the $h_1(1415)\to\phi\eta$ decay, since $\Gamma_{h_1(1415)}$ is very small,  the branching ratio is zero when we take  $n=2$, so we use  $n=3$ to obtain $\mathcal{B}(h_1(1415)\to\phi\eta)$.

Now we relate  the $AVP$ vertex coefficients  $\lambda_{AVP}$ for different decay modes by the SU(3) flavor symmetry/breaking.  For $J^{PC}=1^{++}$
axial-vector mesons decays,  the $AVP$ vertex coefficients can be parameterized as
\begin{eqnarray}
\lambda_{AVP} =c'^A_{0} A_j^iM^j_kM'^k_i+c'^A_{1} A_j^aW^i_aM^j_kM'^k_i+c'^A_{2} A_j^iM^a_kW^j_aM'^k_i+c'^A_{3} A_j^iM^j_kM'^a_iW^k_a, \label{Eq:LambdaAVP}
\end{eqnarray}
where $M^{(')}=V$ or $P$,  $c'^A_{0}$ are the non-perturbative coefficients  under the SU(3) flavor symmetry, and $c'^A_{1,2,3}$  are the non-perturbative SU(3) flavor breaking coefficients.
The vertex coefficients for the $J^{PC}=1^{+-}$ axial-vector mesons decays can be obtained by replacing $A$ to $B$ in Eq. (\ref{Eq:LambdaAVP}).
 To keep the results invariant  under the conjugate change, $c'^{A/B}_1=c'^{A/B}_2$ are used in our analysis.
In terms of Eq. (\ref{Eq:LambdaAVP}), the vertex coefficient relations for the $A/B\to VP$ decays are summarized in Tabs. \ref{Tab:LambdaA2VP1}-\ref{Tab:LambdaA2VP3}.  Although there is not enough phase space for the $K_1(1270)\to \phi K$ decay, we still give its vertex coefficient in Tab.  \ref{Tab:LambdaA2VP1}, because we will use it to obtain one for $K_1(1400)\to \phi K$.
 In Tabs. \ref{Tab:LambdaA2VP1}-\ref{Tab:LambdaA2VP3}, we redefine $F_{1,2,3,4,5}$ and $D_{1,2,3,4,5,6}$ by $c'^A_{0,1,3}$ and $c'^B_{0,1,3}$, respectively. If we ignore the SU(3) flavor breaking effects,
then we have $F_1=F_2=F_3=F_4=F_5= F$ and $D_1=D_2=D_3=D_4=D_5=D_6= D$, where $F=c^A_0$ and $D=c^B_0$.  After ignoring the SU(3) flavor breaking effects,  most of our vertex coefficients  in Tabs. \ref{Tab:LambdaA2VP1}-\ref{Tab:LambdaA2VP3} are in agreement with those in Ref.  \cite{Roca:2003uk}, except for some  with different meson definitions, $i.e.$, $\eta,\eta',h_1,f_1$ mesons. Note that  there is an overall negative  difference for some vertex coefficients due to the different definitions of $D$ and $F$,  and this does not affect the branching ratio results.

\begin{table}[h]
\renewcommand\arraystretch{1.0}
\tabcolsep 0.2in
\centering
\caption{The vertex coefficients  of $K_1(1270)\to VP$ decays.  $F_1=c'^A_0-c'^A_1+c'^A_3,~D_1=c'^B_0-c'^B_1+c'^B_3$, $F_2=c'^A_0-c'^A_1-2c'^A_3,~D_2=c'^B_0-c'^B_1-2c'^B_3$. Ones  of $K_1(1400)\to VP$ decays could be obtained from $K_1(1270)\to VP$ decays by replacing $sin\theta_{K_1} \to cos\theta_{K_1}$ and $cos\theta_{K_1} \to -sin\theta_{K_1}$. }\vspace{0.1cm}
{\footnotesize
\begin{tabular}{lc}  \hline\hline
~~~Decay modes &   Vertex coefficients $\lambda_{AVP}$    \\\hline
$K_1(1270)^-\to K^{*-}\pi^0$&   $\frac{1}{\sqrt{2}}\big(F_1sin\theta_{K_1}+D_1cos\theta_{K_1}\big)$      \\
$K_1(1270)^-\to K^{*-}\eta$&   $\big(F_1sin\theta_{K_1}+D_1cos\theta_{K_1}\big)\big(\frac{cos\theta_P}{\sqrt{6}}-\frac{sin\theta_P}{\sqrt{3}}\big)+\big(F_2sin\theta_{K_1}-D_2cos\theta_{K_1}\big)\big(\frac{2cos\theta_P}{\sqrt{6}}+\frac{sin\theta_P}{\sqrt{3}}\big)$\\
$K_1(1270)^-\to K^{*-}\eta'$&   $\big(F_1sin\theta_{K_1}+D_1cos\theta_{K_1}\big)\big(\frac{sin\theta_P}{\sqrt{6}}+\frac{cos\theta_P}{\sqrt{3}}\big)+\big(F_2sin\theta_{K_1}-D_2cos\theta_{K_1}\big)\big(\frac{2sin\theta_P}{\sqrt{6}}-\frac{cos\theta_P}{\sqrt{3}}\big)$\\
$K_1(1270)^-\to \rho^0K^{-}$&   $\frac{1}{\sqrt{2}}\big(-F_1sin\theta_{K_1}+D_1cos\theta_{K_1}\big)$      \\
$K_1(1270)^-\to \omega K^{-}$&   $\frac{1}{\sqrt{2}}\big(-F_1sin\theta_{K_1}+D_1cos\theta_{K_1}\big)$      \\
$K_1(1270)^-\to \overline{K}^{*0}\pi^-$&   $\big(F_1sin\theta_{K_1}+D_1cos\theta_{K_1}\big)$      \\
$K_1(1270)^-\to \rho^-\overline{K}^{0}$&   $\big(-F_1sin\theta_{K_1}+D_1cos\theta_{K_1}\big)$      \\
$K_1(1270)^-\to \phi K^-$&   $\big(F_2sin\theta_{K_1}+D_2cos\theta_{K_1}\big)$      \\\hline
$\overline{K}_1(1270)^0\to K^{*-}\pi^+$&   $\big(F_1sin\theta_{K_1}+D_1cos\theta_{K_1}\big)$      \\
$\overline{K}_1(1270)^0\to \rho^+K^{-}$&   $\big(-F_1sin\theta_{K_1}+D_1cos\theta_{K_1}\big)$      \\
$\overline{K}_1(1270)^0\to \overline{K}^{*0}\pi^0$&   $-\frac{1}{\sqrt{2}}\big(F_1sin\theta_{K_1}+D_1cos\theta_{K_1}\big)$      \\
$\overline{K}_1(1270)^0\to \overline{K}^{*0}\eta$&   $\big(F_1sin\theta_{K_1}+D_1cos\theta_{K_1}\big)\big(\frac{cos\theta_P}{\sqrt{6}}-\frac{sin\theta_P}{\sqrt{3}}\big)+\big(F_2sin\theta_{K_1}-D_2cos\theta_{K_1}\big)\big(\frac{2cos\theta_P}{\sqrt{6}}+\frac{sin\theta_P}{\sqrt{3}}\big)$\\
$\overline{K}_1(1270)^0\to \overline{K}^{*0}\eta'$&   $\big(F_1sin\theta_{K_1}+D_1cos\theta_{K_1}\big)\big(\frac{sin\theta_P}{\sqrt{6}}+\frac{cos\theta_P}{\sqrt{3}}\big)+\big(F_2sin\theta_{K_1}-D_2cos\theta_{K_1}\big)\big(\frac{2sin\theta_P}{\sqrt{6}}-\frac{cos\theta_P}{\sqrt{3}}\big)$\\
$\overline{K}_1(1270)^0\to \rho^0\overline{K}^{0}$&   $-\frac{1}{\sqrt{2}}\big(-F_1sin\theta_{K_1}+D_1cos\theta_{K_1}\big)$      \\
$\overline{K}_1(1270)^0\to \omega\overline{K}^{0}$&   $\frac{1}{\sqrt{2}}\big(-F_1sin\theta_{K_1}+D_1cos\theta_{K_1}\big)$      \\
$\overline{K}_1(1270)^0\to \phi \overline{K}^0$&   $\big(F_2sin\theta_{K_1}+D_2cos\theta_{K_1}\big)$      \\\hline
\end{tabular}}\label{Tab:LambdaA2VP1}
\renewcommand\arraystretch{1.0}
\tabcolsep 0.2in
\centering
\caption{The vertex coefficients  of the $a_1(1260)/b_1(1235)\to VP$ decays.  $F_3=c'^A_0+2c'^A_1+c'^A_3$, $F_4=c'^A_0+2c'^A_1-2c'^A_3$, $D_3=c'^B_0+2c'^B_1+c'^B_3$, $D_4=c'^B_0+2c'^B_1-2c'^B_3$.  }\vspace{0.1cm}
{\footnotesize
\begin{tabular}{lc|lc}  \hline\hline
~~~Decay modes &   Vertex coefficients $\lambda_{AVP}$     &    ~~~Decay modes                  &       Vertex coefficients $\lambda_{AVP}$    \\\hline
$a_1(1260)^0\to \rho^+\pi^-$                               &       $\sqrt{2}F_3$                &    $b_1(1235)^0\to \rho^0\eta$     &       $2D_3\big(\frac{cos\theta_P}{\sqrt{6}}-\frac{sin\theta_P}{\sqrt{3}}\big)$    \\
$a_1(1260)^0\to \rho^-\pi^+$                               &       $-\sqrt{2}F_3$               &    $b_1(1235)^0\to \rho^0\eta'$    &       $2D_3\big(\frac{sin\theta_P}{\sqrt{6}}+\frac{cos\theta_P}{\sqrt{3}}\big)$     \\
$a_1(1260)^0\to K^{*0}\overline{K}^{0}$                    &       $-\frac{1}{\sqrt{2}}F_4$     &    $b_1(1235)^0\to K^{*0}\overline{K}^{0}$    &       $-\frac{1}{\sqrt{2}}D_4$     \\
$a_1(1260)^0\to \overline{K}^{*0}K^{0}$                    &       $\frac{1}{\sqrt{2}}F_4$      &    $b_1(1235)^0\to \overline{K}^{*0}K^{0}$    &       $-\frac{1}{\sqrt{2}}D_4$     \\
$a_1(1260)^0\to K^{*+}K^{-}$                               &       $\frac{1}{\sqrt{2}}F_4$      &    $b_1(1235)^0\to K^{*+}K^{-}$    &       $\frac{1}{\sqrt{2}}D_4$   \\
$a_1(1260)^0\to K^{*-}K^{+}$                               &       $-\frac{1}{\sqrt{2}}F_4$     &    $b_1(1235)^0\to K^{*-}K^{+}$    &       $\frac{1}{\sqrt{2}}D_4$    \\
$ $                                                        &   $ $                              &    $b_1(1235)^0\to \omega\pi^0$    &       $\sqrt{2}D_3$ \\ \hline
$a_1(1260)^-\to \rho^-\pi^0$                               &       $\sqrt{2}F_3$                &    $b_1(1235)^-\to \rho^-\eta$     &       $2D_3\big(\frac{cos\theta_P}{\sqrt{6}}-\frac{sin\theta_P}{\sqrt{3}}\big)$   \\
$a_1(1260)^-\to \rho^0\pi^-$                               &       $-\sqrt{2}F_3$               &    $b_1(1235)^-\to \rho^-\eta'$    &       $2D_3\big(\frac{sin\theta_P}{\sqrt{6}}+\frac{cos\theta_P}{\sqrt{3}}\big)$    \\
$a_1(1260)^-\to K^{*0}K^{-}$                               &       $F_4$                        &    $b_1(1235)^-\to K^{*0}K^{-}$    &       $D_4$  \\
$a_1(1260)^-\to K^{*-}K^{0}$                               &       $-F_4$                       &    $b_1(1235)^-\to K^{*-}K^{0}$    &       $D_4$     \\
$ $                                                        &      $ $                           &    $b_1(1235)^-\to \omega\pi^-$    &       $\sqrt{2}D_3$ \\ \hline
\end{tabular}}\label{Tab:LambdaA2VP2}
\end{table}
\begin{table}[t]
\renewcommand\arraystretch{1.0}
\tabcolsep 0.3in
\centering
\caption{The vertex coefficients  of the $h_1(1170)/f_1(1285) \to VP$ decays.   $F_5=c'^A_0-4c'^A_1+c'^A_3$, $D_5=c'^B_0-4c'^B_1+c'^B_3$, $D_6=c'^B_0-4c'^B_1-2c'^B_3$.
Ones  of the $h_1(1415)\to VP$ decays could be obtained from the $h_1(1170)\to VP$ decays by replacing $sin\theta_{1P_1} \to cos\theta_{1P_1}$ and $cos\theta_{1P_1} \to -sin\theta_{1P_1}$.
Ones  of the $f_1(1420)\to VP$ decays could be obtained from the $f_1(1285)\to VP$ decays by replacing $sin\theta_{3P_1} \to cos\theta_{3P_1}$ and $cos\theta_{3P_1} \to -sin\theta_{3P_1}$.}\vspace{0.02cm}
{\footnotesize
\begin{tabular}{lclc}  \hline\hline
~~~Decay modes                          &          Vertex coefficients $\lambda_{AVP}$                                                                                   
\\\hline
$h_1(1170)\to \rho^0\pi^0$            &          $\frac{2} {\sqrt{6}}D_3\big(\sqrt{2}cos\theta_{1P_1}+sin\theta_{1P_1}\big)$                                             
\\
$h_1(1170)\to \omega\eta$             &          $\frac{\sqrt{2}}{3}D_3\big(\sqrt{2}cos\theta_{1P_1}+sin\theta_{1P_1}\big)\big(cos\theta_P-\sqrt{2}sin\theta_P\big)$     
\\
$h_1(1170)\to \omega\eta'$            &          $\frac{\sqrt{2}}{3}D_3\big(\sqrt{2}cos\theta_{1P_1}+sin\theta_{1P_1}\big)\big(sin\theta_P+\sqrt{2}cos\theta_P\big)$     
\\
$h_1(1170)\to \rho^+\pi^-$            &          $\frac{2}{\sqrt{6}}D_3\big(\sqrt{2}cos\theta_{1P_1}+sin\theta_{1P_1}\big)$                                              
\\
$h_1(1170)\to \rho^-\pi^+$            &          $\frac{2}{\sqrt{6}}D_3\big(\sqrt{2}cos\theta_{1P_1}+sin\theta_{1P_1}\big)$                                              
\\
$h_1(1170)\to K^{*+}K^-$              &          $ \frac{cos\theta_{1P_1}}{\sqrt{3}}\big(D_4+D_5\big)+\frac{sin\theta_{1P_1}}{\sqrt{6}}\big(D_4-2D_5\big)$               
\\
$h_1(1170)\to K^{*-}K^+$              &          $ \frac{cos\theta_{1P_1}}{\sqrt{3}}\big(D_4+D_5\big)+\frac{sin\theta_{1P_1}}{\sqrt{6}}\big(D_4-2D_5\big)$               
\\
$h_1(1170)\to K^{*0}\overline{K}^0$   &          $ \frac{cos\theta_{1P_1}}{\sqrt{3}}\big(D_4+D_5\big)+\frac{sin\theta_{1P_1}}{\sqrt{6}}\big(D_4-2D_5\big)$               
\\
$h_1(1170)\to \overline{K}^{*0}K^0$   &          $ \frac{cos\theta_{1P_1}}{\sqrt{3}}\big(D_4+D_5\big)+\frac{sin\theta_{1P_1}}{\sqrt{6}}\big(D_4-2D_5\big)$               
\\
$h_1(1170)\to \phi\eta$               &          $\frac{2}{3}D_6\big(-cos\theta_{1P_1}+\sqrt{2}sin\theta_{1P_1}\big)\big(\sqrt{2}cos\theta_{P}+sin\theta_{P}\big)$       
\\
$h_1(1170)\to \phi\eta'$              &          $\frac{2}{3}D_6\big(-cos\theta_{1P_1}+\sqrt{2}sin\theta_{1P_1}\big)\big(\sqrt{2}sin\theta_{P}-cos\theta_{P}\big)$       
\\\hline
$f_1(1285)\to K^{*+}K^-$              &          $ \frac{cos\theta_{3P_1}}{\sqrt{3}}\big(F_4-F_5\big)+\frac{sin\theta_{3P_1}}{\sqrt{6}}\big(F_4+2F_5\big)$               
\\
$f_1(1285)\to K^{*-}K^+$              &          $ -\frac{cos\theta_{3P_1}}{\sqrt{3}}\big(F_4-F_5\big)-\frac{sin\theta_{3P_1}}{\sqrt{6}}\big(F_4+2F_5\big)$              
\\
$f_1(1285)\to K^{*0}\overline{K}^0$   &          $ \frac{cos\theta_{3P_1}}{\sqrt{3}}\big(F_4-F_5\big)+\frac{sin\theta_{3P_1}}{\sqrt{6}}\big(F_4+2F_5\big)$               
\\
$f_1(1285)\to \overline{K}^{*0}K^0$   &          $ -\frac{cos\theta_{3P_1}}{\sqrt{3}}\big(F_4-F_5\big)-\frac{sin\theta_{3P_1}}{\sqrt{6}}\big(F_4+2F_5\big)$              
\\\hline
\end{tabular}}\label{Tab:LambdaA2VP3}
\end{table}

\subsubsection{Numerical results for the $A\to VP$ decays} \label{sec:A2VP}

In the $A\to VP$ decays, only a few  branching ratios of $K_1 (1270)/K_1 (1400)\to VP $  have been measured  \cite{PDG2022}
{\small
\begin{eqnarray}
&&\mathcal{B}(K_1 (1270)\to \rho K)=(38\pm13)\%,~~~~\mathcal{B}(K_1 (1270)\to \omega K)=(11\pm2)\%, ~~~~\mathcal{B}(K_1 (1270)\to K^*\pi)=(21\pm10)\%, \label{Eq:K12702VPdata}\\
&&\mathcal{B}(K_1 (1400)\to \omega K)=(1\pm1)\%,~~~~~~~\mathcal{B}(K_1 (1400)\to \rho K)=(3\pm3)\%,~~~~~~\mathcal{B}(K_1 (1400)\to K^*\pi)=(94\pm6)\%. \label{Eq:K14002VPdata}
\end{eqnarray}}
In addition, the experimental errors of $\mathcal{B}(K_1 (1400)\to \rho K)$, $\mathcal{B}(K_1 (1400)\to \omega K)$ and $\mathcal{B}(K_1 (1270)\to K^*(892)\pi)$ are quite large.
Given the insufficient  experimental data and consistency with our study of semileptonic decay $D\to A\ell^+\nu_\ell$, we only consider the SU(3) flavor symmetry for the  $A\to VP$ numerical results.

Some values of  $D$ and $F$ have been obtained  from $A\to VP$ decays \cite{Roca:2003uk}.  Considering the results in Ref. \cite{Roca:2003uk},  we conservatively choose $F\in[0.6, 2.0]$ and $D\in[-1.6, -0.4]$ as the   original ranges.
Then, using the experimental data of $\mathcal{B}(K_1\to VP)$ in Eqs. (\ref{Eq:K12702VPdata}-\ref{Eq:K14002VPdata}),  $\mathcal{B}(a_1(1260)/b_1(1235)/h_1(1170)/$ $h_1(1415)/f_1(1285)/f_1(1420)\to VP)\leq1$, and also considering an extra constraint from   $\mathcal{B}(D\to K_1(1270)e^+\nu_e)$  in Eqs. (\ref{Eq:D2K1lvdata1}-\ref{Eq:D2K1lvdata2}) and $\mathcal{B}(D\to Ae^+\nu_e,A\to PV)$ in Eqs. (\ref{Eq:BrB2PVlvA1}-\ref{Eq:BrB2PVlvA2}), one can constrain the relevant parameters $F$, $D$, and the mixing angles $\theta_{K_1}$,  $\theta_{3P_1}$, and $\theta_{1P_1}$. Three constrained mixing angles are given in Eqs. (\ref{Eq:theta1}-\ref{Eq:theta2}), and the constrained $F$ and $D$ are
\begin{eqnarray}
&&F\in[0.94, 1.78],~~~~~~D\in[-1.30, -0.64]~~~~\mbox{within}~~~2\sigma ~~~\mbox{errors},\label{Eq:FD1}\\
&&F\in[1.42, 1.66],~~~~~~D\in[-1.21, -0.87]~~~~\mbox{within}~~~1\sigma ~~~\mbox{errors}.\label{Eq:FD2}
\end{eqnarray}
Using the relevant  constrained  parameters, we  give the predictions of $\mathcal{B}(A\to VP)$,  which have not been measured to date.  The numerical results are given in Tab.  \ref{Tab:AB2VP}.
\begin{table}[hbt]
\renewcommand\arraystretch{1.0}
\tabcolsep 0.25in
\centering
\caption{The branching ratio predictions  of the $K_1/a_1/b_1/f_1/h_1\to VP$ decays by the SU(3)  flavor symmetry with $1\sigma$ error.    The unit is $10^{-2}$ for all branching ratios.   }\vspace{0.1cm}
{\scriptsize
\begin{tabular}{lc|lc}  \hline\hline
Branching ratios                                                      &  Our predictions                   &                                                                                      & Our predictions        \\\hline
$\mathcal{B}(K_1(1270)^-\to K^{*-}\pi^0)$                               & $7.24\pm3.17$                    &               $\mathcal{B}(K_1(1400)^-\to K^{*-}\pi^0)$                               &  $30.54\pm1.01$              \\
$\mathcal{B}(K_1(1270)^-\to K^{*-}\eta)$                                &  $0.50\pm0.16$                   &               $\mathcal{B}(K_1(1400)^-\to K^{*-}\eta)$                                &  $2.81\pm1.84$              \\
$\mathcal{B}(K_1(1270)^-\to \rho^0K^{-})$                               &  $16.82\pm0.49 $                 &               $\mathcal{B}(K_1(1400)^-\to \rho^0K^{-})$                               &  $0.75\pm0.75$              \\
$\mathcal{B}(K_1(1270)^-\to \omega K^{-})$                              &  $9.52\pm0.21$                   &               $\mathcal{B}(K_1(1400)^-\to \omega K^{-})$                              &  $0.73\pm0.73$              \\
$\mathcal{B}(K_1(1270)^-\to \overline{K}^{*0}\pi^-)$                    &  $14.20\pm6.22$                   &              $\mathcal{B}(K_1(1400)^-\to \overline{K}^{*0}\pi^-)$                    &  $60.32\pm1.97$             \\
$\mathcal{B}(K_1(1270)^-\to \rho^-\overline{K}^{0})$                    &  $32.71\pm0.82$                   &              $\mathcal{B}(K_1(1400)^-\to \rho^-\overline{K}^{0})$                    &  $1.48\pm1.48$              \\
%
$\mathcal{B}(\overline{K}_1(1270)^0\to K^{*-}\pi^+)$                    &  $14.39\pm6.31$                   &                $\mathcal{B}(K_1(1400)^-\to \phi K^-)$                                  &  $3.81\pm0.65$                      \\
$\mathcal{B}(\overline{K}_1(1270)^0\to \rho^+K^{-})$                    &  $33.91\pm0.86$                  &                  $\mathcal{B}(\overline{K}_1(1400)^0\to K^{*-}\pi^+)$                    &  $60.88\pm2.01$             \\
$\mathcal{B}(\overline{K}_1(1270)^0\to \overline{K}^{*0}\pi^0)$         &  $7.15\pm3.13$                   &                  $\mathcal{B}(\overline{K}_1(1400)^0\to \rho^+K^{-})$                    &  $1.51\pm1.51$               \\
$\mathcal{B}(\overline{K}_1(1270)^0\to \overline{K}^{*0}\eta)$          &  $0.41\pm0.14$                   &                  $\mathcal{B}(\overline{K}_1(1400)^0\to \overline{K}^{*0}\pi^0)$         &  $30.27\pm0.99$              \\
$\mathcal{B}(\overline{K}_1(1270)^0\to \rho^0\overline{K}^{0})$         &  $16.22\pm0.47$                  &                  $\mathcal{B}(\overline{K}_1(1400)^0\to \overline{K}^{*0}\eta)$          &  $2.67\pm1.75$              \\
$\mathcal{B}(\overline{K}_1(1270)^0\to \omega\overline{K}^{0})$         &  $8.87\pm0.20$                   &                  $\mathcal{B}(\overline{K}_1(1400)^0\to \rho^0\overline{K}^{0})$         &  $0.74\pm0.74$               \\
%
$\mathcal{B}(K_1(1270)\to K^*\pi)$                                      &   $21.48\pm9.42$                      &             $\mathcal{B}(\overline{K}_1(1400)^0\to \omega\overline{K}^{0})$         &  $0.71\pm0.71$               \\
$\mathcal{B}(K_1(1270)\to \rho K)$                                      &  $49.78\pm1.21$                        &              $\mathcal{B}(\overline{K}_1(1400)^0\to \phi\overline{K}^{0})$           &  $3.64\pm0.63$              \\
$\mathcal{B}(K_1(1270)\to \omega K)$                                    &  $9.20\pm0.20$                         &              $\mathcal{B}(K_1(1400)\to K^*\pi)$                                      &  $91.01\pm2.99$                 \\
$\mathcal{B}(K_1(1270)\to K^*\eta )$                                    &  $0.45\pm0.15$                    &                   $\mathcal{B}(K_1(1400)\to \rho K)$                                      &  $2.24\pm2.24$                  \\
                                                                        &                                 &                      $\mathcal{B}(K_1(1400)\to \omega K)$                                    &  $0.72\pm0.72$                  \\
                                                                        &                                                 &       $\mathcal{B}(K_1(1400)\to K^*\eta )$                                    &  $2.74\pm1.79$                      \\
                                                                        &
                                                                                                                          &      $\mathcal{B}(K_1(1400)\to \phi K)$                                       &  $3.73\pm0.64$                      \\ \hline
%
$\mathcal{B}(a_1(1260)^0\to \rho^+\pi^-)$              &  $ 38.15\pm8.46$                     & $\mathcal{B}(b_1(1235)^0\to \rho^0\eta)$               &    $5.07\pm2.14 $     \\
$\mathcal{B}(a_1(1260)^0\to \rho^-\pi^+)$              &  $38.15\pm8.46$                      & $\mathcal{B}(b_1(1235)^0\to \omega\pi^0)$              &  $65.46\pm22.46$      \\
$\mathcal{B}(a_1(1260)^0\to K^{*0}\overline{K}^{0})$   &  $1.82\pm0.52$                       & $\mathcal{B}(b_1(1235)^0\to K^{*0}\overline{K}^{0})$   &  $0.37\pm0.15$        \\
$\mathcal{B}(a_1(1260)^0\to \overline{K}^{*0}K^{0})$   & $1.82\pm0.52$                        & $\mathcal{B}(b_1(1235)^0\to \overline{K}^{*0}K^{0})$   &  $0.37\pm0.15$        \\
$\mathcal{B}(a_1(1260)^0\to K^{*+}K^{-})$              &   $1.89\pm0.53$                      & $\mathcal{B}(b_1(1235)^0\to K^{*+}K^{-})$              &  $ 0.42\pm0.17$       \\
$\mathcal{B}(a_1(1260)^0\to K^{*-}K^{+})$              &   $1.89\pm0.53$                      & $\mathcal{B}(b_1(1235)^0\to K^{*-}K^{+})$              &  $ 0.42\pm0.17$       \\
$\mathcal{B}(a_1(1260)^-\to \rho^-\pi^0)$              &     $38.33\pm8.51 $                  & $\mathcal{B}(b_1(1235)^-\to \rho^-\eta)$               & $5.13\pm2.16$         \\
$\mathcal{B}(a_1(1260)^-\to \rho^0\pi^-)$              &   $38.13\pm8.45$                     & $\mathcal{B}(b_1(1235)^-\to \omega\pi^-)$              &  $65.17\pm22.36 $     \\
$\mathcal{B}(a_1(1260)^-\to K^{*0}K^{-})$              &   $ 3.70\pm1.04 $                    & $\mathcal{B}(b_1(1235)^-\to K^{*0}K^{-})$              &  $0.78\pm0.32$        \\
$\mathcal{B}(a_1(1260)^-\to K^{*-}K^{0})$              &    $3.72\pm1.04$                     & $\mathcal{B}(b_1(1235)^-\to K^{*-}K^{0})$              &  $0.79\pm0.33$        \\
$\mathcal{B}(a_1(1260)\to\rho\pi)$                     &  $76.38\pm16.93$                     & $\mathcal{B}(b_1(1235)\to\rho\eta)$                    &   $5.10\pm2.15 $      \\
$\mathcal{B}(a_1(1260)\to K^{*}K)$                     &  $7.42\pm2.09  $                     & $\mathcal{B}(b_1(1235)\to\omega\pi)$                   &   $65.31\pm22.41$     \\
                                                       &                                       & $\mathcal{B}(b_1(1235)\to K^{*}K)$                     &   $1.58\pm0.65$       \\\hline
$\mathcal{B}(f_1(1285)\to K^{*+}K^-)$                & $0.62\pm0.24$          &$\mathcal{B}(f_1(1420)\to K^{*+}K^-)$                & $13.09\pm13.09$                   \\
$\mathcal{B}(f_1(1285)\to K^{*-}K^+)$                & $0.62\pm0.24$          &$\mathcal{B}(f_1(1420)\to K^{*-}K^+)$                & $13.09\pm13.09$                 \\
$\mathcal{B}(f_1(1285)\to K^{*0}\overline{K}^0)$     & $0.13\pm0.05$          &$\mathcal{B}(f_1(1420)\to K^{*0}\overline{K}^0)$     & $11.75\pm11.75$                 \\
$\mathcal{B}(f_1(1285)\to \overline{K}^{*0}K^0)$     & $0.13\pm0.05$          &$\mathcal{B}(f_1(1420)\to \overline{K}^{*0}K^0)$     & $11.75\pm11.75$                 \\
$\mathcal{B}(f_1(1285)\to {K}K^{*})$                 & $1.49\pm0.59 $         &$\mathcal{B}(f_1(1420)\to K^{*}K)$                 & $49.64\pm49.64 $                   \\\hline
$\mathcal{B}(h_1(1170)\to \rho^0\pi^0)$                &   $21.84\pm9.33$       &$\mathcal{B}(h_1(1415)\to \rho^0\pi^0)$                &   $13.20\pm13.20$                       \\
$\mathcal{B}(h_1(1170)\to \omega\eta)$                 &   $2.01\pm1.04$        &$\mathcal{B}(h_1(1415)\to \omega\eta)$                 &   $3.37\pm3.37$                         \\
$\mathcal{B}(h_1(1170)\to \rho^+\pi^-)$                &   $21.74\pm9.29$       &$\mathcal{B}(h_1(1415)\to \rho^+\pi^-)$                &  $13.17\pm13.17$                        \\
$\mathcal{B}(h_1(1170)\to \rho^-\pi^+)$                &   $21.74\pm9.29$       &$\mathcal{B}(h_1(1415)\to \rho^-\pi^+)$                &   $13.17\pm13.17$                       \\
$\mathcal{B}(h_1(1170)\to K^{*+}K^-)$                  &   $0.85\pm0.76$        &$\mathcal{B}(h_1(1415)\to K^{*+}K^-)$                  &   $12.15\pm9.79$                       \\
$\mathcal{B}(h_1(1170)\to K^{*-}K^+)$                  &   $0.85\pm0.76$        &$\mathcal{B}(h_1(1415)\to K^{*-}K^+)$                  &   $12.15\pm9.79$                       \\
$\mathcal{B}(h_1(1170)\to K^{*0}\overline{K}^0)$       &   $0.81\pm0.73$        &$\mathcal{B}(h_1(1415)\to K^{*0}\overline{K}^0)$       &   $10.42\pm8.50$                        \\
$\mathcal{B}(h_1(1170)\to \overline{K}^{*0}K^0)$       &   $0.81\pm0.73$        &$\mathcal{B}(h_1(1415)\to \overline{K}^{*0}K^0)$       &   $10.42\pm8.50$                        \\
$\mathcal{B}(h_1(1170)\to \phi\eta)$                   &   $0.08\pm0.08$        &$\mathcal{B}(h_1(1415)\to \phi\eta)$                   &   $1.19\pm0.84$                   \\
$\mathcal{B}(h_1(1170)\to \rho\pi)$                    &   $65.31\pm27.91$      &$\mathcal{B}(h_1(1415)\to \rho\pi)$                    &   $39.54\pm39.53$                       \\
$\mathcal{B}(h_1(1170)\to K^{*}K)$                     &   $3.33\pm2.98$        &$\mathcal{B}(h_1(1415)\to K^{*}K)$                     &   $45.13\pm36.58$                         \\\hline
\end{tabular}}\label{Tab:AB2VP}
\end{table}
Most of our branching ratio predictions have large errors, since we consider the errors of the decay widths and the masses;  we conservatively choose the  parameters $D$ and $F$ as well as three mixing angles,  and consider the present large experimental errors of $K_1 (1270)/K_1 (1400)\to VP $.  If the result is shown as $a\pm a$, it means that the allowed lower limit is very small, and only the upper limit $2a$ is obtained.

Fitted results were previously  obtained by three possible solutions in Ref. \cite{Roca:2003uk}.
Apart from the measured decays of $K_1(1270)$ and $K_1(1400)$ mesons given in Eqs. (\ref{Eq:K12702VPdata}-\ref{Eq:K14002VPdata}),  the average values  of the three  fitted central values are  \cite{Roca:2003uk}
\begin{eqnarray}
&&\mathcal{B}(a_1(1260)\to\rho\pi) = 62.75\%, ~~~~~~~~~~~~~~~~~\mathcal{B}(a_1(1260)\to K^{*}K)= 7.53\%, \\
&&\mathcal{B}(b_1(1235)\to\omega\pi)=73.47\%,~~~~~~~~~~~~~~~~~\mathcal{B}(b_1(1235)\to K^{*}K)=5.92\%.
\end{eqnarray}
Our prediction  of   $\mathcal{B}(b_1(1235)\to K^{*}K)$ is smaller than that in Ref. \cite{Roca:2003uk}; nevertheless, the other three branching ratios are consistent with each other.

\begin{figure}[hb]
\begin{center}
\includegraphics[scale=0.55]{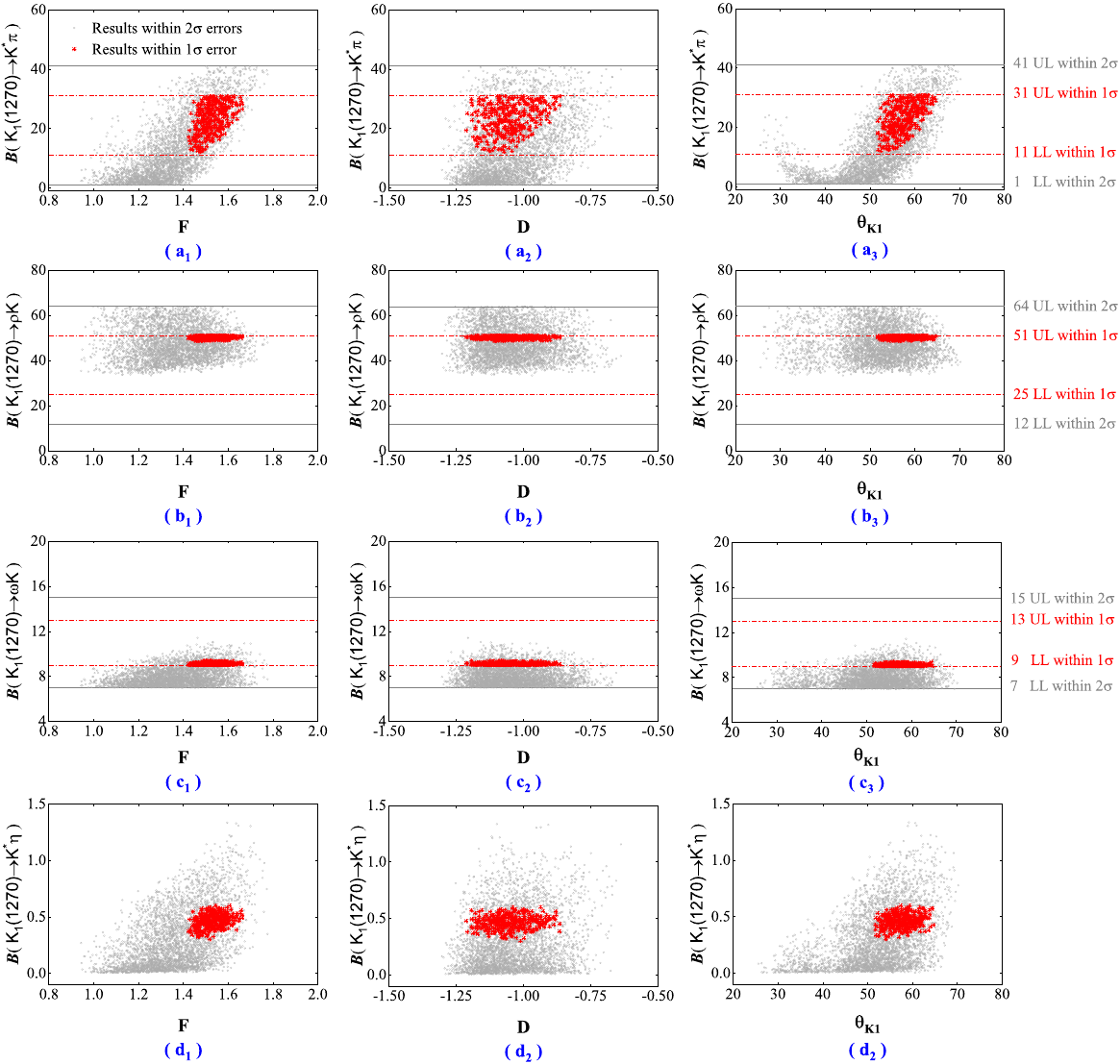}
\end{center}
\caption{The constrained effects of the non-perturbative parameters in the $K_1(1270)\to VP$ decays (in units of $10^{-2}$).  }\label{fig:A2VP1}
\end{figure}

\begin{figure}[htb]
\begin{center}
\includegraphics[scale=0.46]{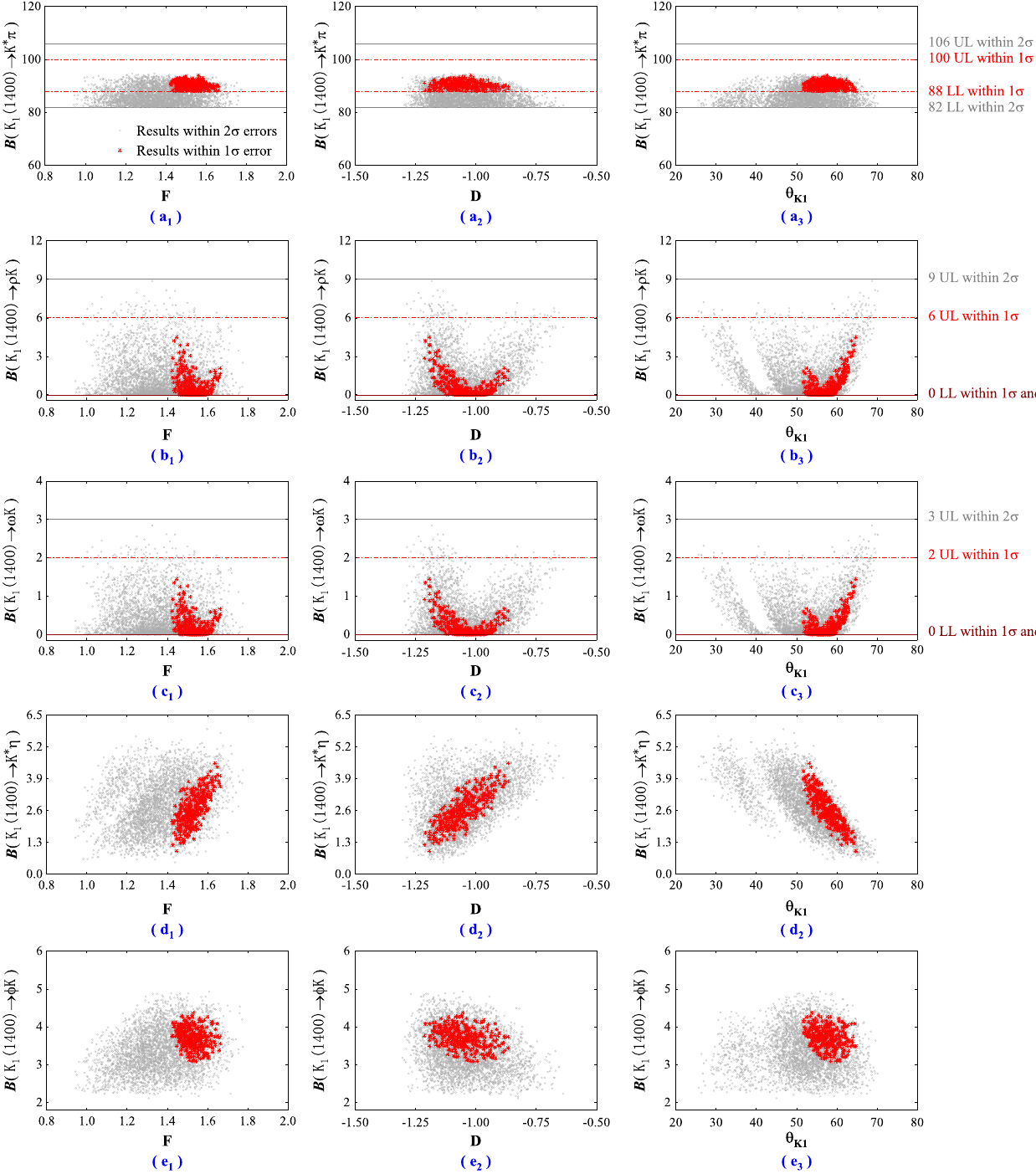}
\end{center}
\caption{The constrained effects of the non-perturbative parameters in the $K_1(1400)\to VP$ decays (in units of $10^{-2}$).  }\label{fig:A2VP2}
%
\begin{center}
\includegraphics[scale=0.46]{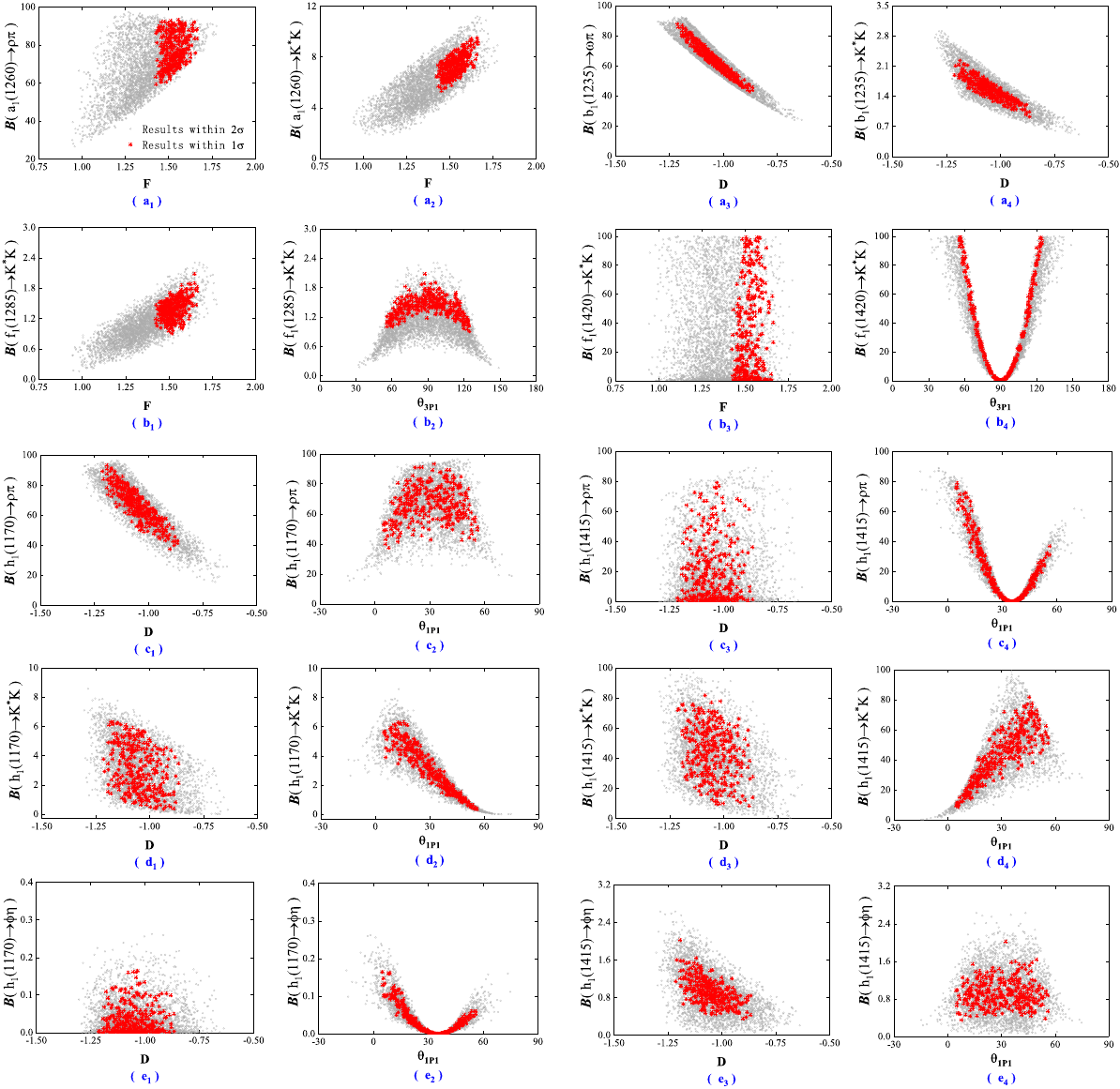}
\end{center}
\caption{The constrained effects of the non-perturbative parameters in the $a_1/b_1/f_1/h_1\to VP$ decays ( in units of $10^{-2}$). }\label{fig:A2VP3}
\end{figure}

The sensitivity of  $\mathcal{B}(A \to VP)$ to the constrained parameters is shown in Figs. \ref{fig:A2VP1}-\ref{fig:A2VP3}.
As shown in Figs. \ref{fig:A2VP1}-\ref{fig:A2VP2}, the upper limits of $K_1(1270)\to K^*\pi,\rho K$ and $K_1(1400)\to \rho K$ and the lower limits of $K_1(1270)\to K^*\pi, \omega K$ and $K_1(1400)\to K^*\pi,\rho K,\omega K$ give the effective constraints on the relevant non-perturbative parameters.  $\mathcal{B}(K_1(1270)\to K^*\pi)$  and  $\mathcal{B}(K_1(1270)\to K^*\eta)$ are sensitive to  $F$ and $\theta_{K1}$.
$\mathcal{B}(K_1(1400)\to \rho K)$, $\mathcal{B}(K_1(1400)\to  \omega K)$   and $\mathcal{B}(K_1(1400)\to K^*\eta)$  are sensitive to  $D$ and $\theta_{K1}$.
$\mathcal{B}(K_1(1400)\to \rho K)$ and $\mathcal{B}(K_1(1400)\to  \omega K)$  have similar change trends, and the  inflection point values appear when $D\approx-1.0$ and $\theta_{K_1}\approx56^\circ$.
As displayed  in Fig. \ref{fig:A2VP3},  many $\mathcal{B}(a_1/b_1/f_1/h_1\to VP)$ are very sensitive to  $F$, $D$,  $\theta_{3_{P1}}$ and $\theta_{1_{P1}}$.  Any  future measurement of these decays  can obviously place constraints on the relevant non-perturbative parameters.


\clearpage

\subsection{Results of the $D\to A(A\to VP)\ell^+\nu$ decays } \label{sec:D2A2VPlv}
For the four-body semileptonic decays with resonances,  in the case where the decay widths of
the resonances are very narrow, the resonant decay branching ratios respect a simple factorization relation
\begin{eqnarray}
\mathcal{B}(D\to R\ell^+\nu_\ell,R\to VP)=\mathcal{B}(D\to R\ell^+\nu_\ell)\times\mathcal{B}(R\to VP), \label{Eq:BrD2AlvA2VP}
\end{eqnarray}
where $R$ denotes the resonance. This relation is also a good approximation for wider resonances.

At present, only  experimental upper limits of $\mathcal{B}(D^+\to b_1(1235)^0e^+\nu_e,~b_1(1235)^0\to \omega\pi^0)$ and $\mathcal{B}(D^0\to b_1(1235)^-e^+\nu_e,~b_1(1235)^-\to \omega\pi^-)$  have been reported \cite{PDG2022,BESIII:2020jan}
\begin{eqnarray}
\mathcal{B}(D^+\to b_1(1235)^0e^+\nu_e,~b_1(1235)^0\to \omega\pi^0)\leq 1.75\times 10^{-4}, \label{Eq:BrB2PVlvA1} \\
\mathcal{B}(D^0\to b_1(1235)^-e^+\nu_e,~b_1(1235)^-\to \omega\pi^-)\leq 1.12 \times 10^{-4}. \label{Eq:BrB2PVlvA2}
\end{eqnarray}
Branching ratios of $D\to A(A\to VP)\ell^+\nu_\ell$ decays can be obtained by using $\mathcal{B}(D\rightarrow A\ell^+ \nu_\ell)$ given in Sec. \ref{Sec:D2Alv} and $\mathcal{B}(A\rightarrow VP)$ given in Sec. \ref{Sec:A2VP} as well as considering the further constraints in Eqs. (\ref{Eq:BrB2PVlvA1}-\ref{Eq:BrB2PVlvA2}), and our predictions  are listed in the second and fourth columns of  Tabs. \ref{Tab:D2PVlvAT1}-\ref{Tab:D2PVlvAT3}.

\begin{table}[b]
\renewcommand\arraystretch{1}
\tabcolsep 0.08in
\centering
\caption{SU(3) Branching ratios of  $D^0\to VP\ell^+\nu_{\ell}(\ell=e/\mu )$ decays within 1$\sigma$ error. The unit is $10^{-5}$ for the branching ratios. $\mathcal{B}_{[A]}$ and $\mathcal{B}_{[T]}$ denote the axial meson resonance and tensor meson resonance, respectively.} \vspace{0.08cm}
{\footnotesize
\begin{tabular}{l|cc|cc}  \hline\hline
Decay modes & $\mathcal{B}_{[A]}(\ell=e)$ & $\mathcal{B}_{[T]}(\ell=e)$ & $\mathcal{B}_{[A]}(\ell=\mu)$  & $\mathcal{B}_{[T]}(\ell=\mu)$ \\\hline
$D^0\to K^{*-}\pi^0\ell^+\nu_\ell$  &$7.29\pm3.48_{[K_1(1270)^-]}$ &$0.11\pm0.05_{[K^*_2(1430)^-]}$ &$6.29\pm3.00_{[K_1(1270)^-]}$    &$0.08\pm0.03_{[K^*_2(1430)^-]} $ \\
& $0.60\pm0.60_{[K_1(1400)^-]}$ &  &$0.47\pm0.47_{[K_1(1400)^-]}$   & \\\hline
$D^0\to \overline{K}^{*0}\pi^-\ell^+\nu_\ell$  &$14.30\pm6.83_{[K_1(1270)^-]}$&$0.22\pm0.09_{[K^*_2(1430)^-]}$ &$12.33\pm5.89_{[K_1(1270)^-]}$    &$0.15\pm0.05_{[K^*_2(1430)^-]} $ \\
 & $1.19\pm1.19_{[K_1(1400)^-]}$ &  &$0.93\pm0.93_{[K_1(1400)^-]}$   & \\\hline
$D^0\to \rho^0 K^-\ell^+\nu_\ell$  &$15.77\pm1.97_{[K_1(1270)^-]}$&$0.04\pm0.02_{[K^*_2(1430)^-]}$ &$13.61\pm1.70_{[K_1(1270)^-]}$    &$0.03\pm0.01_{[K^*_2(1430)^-]} $ \\
 & $0.03\pm0.03_{[K_1(1400)^-]}$ &  &$0.02\pm0.02_{[K_1(1400)^-]}$   & \\\hline
$D^0\to \rho^- \overline{K}^0\ell^+\nu_\ell$  &$30.64\pm3.88_{[K_1(1270)^-]}$&$0.08\pm0.03_{[K^*_2(1430)^-]}$ &$26.44\pm3.36_{[K_1(1270)^-]}$    &$0.05\pm0.02_{[K^*_2(1430)^-]} $ \\
 & $0.06\pm0.06_{[K_1(1400)^-]}$ &  &$0.05\pm0.05_{[K_1(1400)^-]}$   & \\\hline
$D^0\to \omega K^-\ell^+\nu_\ell$ &$8.93\pm1.12_{[K_1(1270)^-]}$&$0.04\pm0.02_{[K^*_2(1430)^-]}$ &$7.70\pm0.96_{[K_1(1270)^-]}$    &$0.03\pm0.02_{[K^*_2(1430)^-]} $ \\
 & $0.03\pm0.03_{[K_1(1400)^-]}$ &  &$0.02\pm0.02_{[K_1(1400)^-]}$   & \\\hline
$D^0\to K^{*-}\eta \ell^+\nu_\ell$  &$0.48\pm0.20_{[K_1(1270)^-]}$&$0.008\pm0.003_{[K^*_2(1430)^-]}$ &$0.41\pm0.17_{[K_1(1270)^-]}$    &$0.005\pm0.002_{[K^*_2(1430)^-]} $ \\&
$0.04\pm0.04_{[K_1(1400)^-]}$ &  &$0.03\pm0.03_{[K_1(1400)^-]}$   & \\\hline
$D^0\to \phi K^-\ell^+\nu_\ell$  &$0.07\pm0.07_{[K_1(1400)^-]}$&$0.0005\pm0.0002_{[K^*_2(1430)^-]}$&$0.06\pm0.06_{[K_1(1400)^-]}$ &$0.0004\pm0.0002_{[K^*_2(1430)^-]} $ \\\hline
$D^0\to \rho^-\pi^0\ell^+\nu_\ell$  &$1.87\pm1.09_{[a_1(1260)^-]}$&$0.12\pm0.05_{[a_2(1320)^-]}$ &$1.64\pm0.97_{[a_1(1260)^-]}$    &$0.09\pm0.03_{[a_2(1320)^-]} $ \\\hline
$D^0\to \rho^0\pi^-\ell^+\nu_\ell$ &$1.87\pm1.08_{[a_1(1260)^-]}$&$0.12\pm0.05_{[a_2(1320)^-]}$ &$1.64\pm0.96_{[a_1(1260)^-]}$    &$0.09\pm0.03_{[a_2(1320)^-]} $ \\\hline
$D^0\to K^{*0}K^-\ell^+\nu_\ell$ &$0.15\pm0.07_{[a_1(1260)^-]}$&$0.0003\pm0.0001_{[a_2(1320)^-]}$ &$0.13\pm0.06{[a_1(1260)^-]}$   &$0.0003\pm0.0001_{[a_2(1320)^-]} $ \\
& $0.02\pm0.02_{[b_1(1235)^-]}$  & & $0.02\pm0.01_{[b_1(1235)^-]}$ &   \\\hline
$D^0\to K^{*-}K^0\ell^+\nu_\ell$  &$0.15\pm0.07_{[a_1(1260)^-]}$&$0.0004\pm0.0002_{[a_2(1320)^-]}$ &$0.13\pm0.06_{[a_1(1260)^-]}$   &$0.0003\pm0.0001_{[a_2(1320)^-]} $ \\
 & $0.02\pm0.02_{[b_1(1235)^-]}$  & & $0.02\pm0.02_{[b_1(1235)^-]}$ &   \\\hline
$D^0\to \omega\pi^-\ell^+\nu_\ell$  &$1.89\pm1.26_{[b_1(1235)^-]}$ & &$1.65\pm1.11_{[b_1(1235)^-]}$&   \\\hline
$D^0\to \rho^-\eta \ell^+\nu_\ell$  &$0.16\pm0.10_{[b_1(1235)^-]}$ & &$0.14\pm0.09_{[b_1(1235)^-]}$&   \\\hline
\end{tabular}}\label{Tab:D2PVlvAT1}
\end{table}

\begin{table}[h]
\renewcommand\arraystretch{0.8}
\tabcolsep 0.05in
\centering
\caption{SU(3) Branching ratios of  $D^+\to VP\ell^+\nu_{\ell}(\ell=e/\mu )$ decays within 1$\sigma$ error. The unit is $10^{-5}$ for the branching ratios. $\mathcal{B}_{[A]}$ and $\mathcal{B}_{[T]}$ denote the axial meson resonance and tensor meson resonance respectively.}\vspace{0.10cm}
{\footnotesize
\begin{tabular}{l|cc|cc}  \hline\hline
Decay modes & $\mathcal{B}_{[A]}(\ell=e)$  & $\mathcal{B}_{[T]}(\ell=e)$  & $\mathcal{B}_{[A]}(\ell=\mu)$ & $\mathcal{B}_{[T]}(\ell=\mu)$ \\\hline
$D^+\to K^{*-}\pi^+\ell^+\nu_\ell$  &$37.70\pm17.93_{[\overline{K}_1(1270)^0]}$&$0.59\pm0.23_{[\overline{K}^*_2(1430)^0]}$&$32.59\pm15.51_{[\overline{K}_1(1270)^0]}$ &$0.39\pm0.14_{[\overline{K}^*_2(1430)^0]} $ \\
& $3.17\pm3.17_{[\overline{K}_1(1400)^0]}$  &  & $2.49\pm2.49_{[\overline{K}_1(1400)^0]}$ &   \\\hline
$D^+\to \overline{K}^{*0}\pi^0\ell^+\nu_\ell$  &$18.73\pm8.91_{[\overline{K}_1(1270)^0]}$&$0.29\pm0.11_{[\overline{K}^*_2(1430)^0]}$&$16.19\pm7.71_{[\overline{K}_1(1270)^0]}$ &$0.19\pm0.07_{[\overline{K}^*_2(1430)^0]} $ \\
& $1.58\pm1.58_{[\overline{K}_1(1400)^0]}$  &  & $1.24\pm1.24_{[\overline{K}_1(1400)^0]}$ &   \\\hline
$D^+\to \rho^+ K^-\ell^+\nu_\ell$  &$82.47\pm10.58_{[\overline{K}_1(1270)^0]}$&$0.21\pm0.09_{[\overline{K}^*_2(1430)^0]}$&$71.30\pm9.16_{[\overline{K}_1(1270)^0]}$ &$0.14\pm0.06_{[\overline{K}^*_2(1430)^0]} $ \\
& $0.16\pm0.16_{[\overline{K}_1(1400)^0]}$  &  & $0.13\pm0.13_{[\overline{K}_1(1400)^0]}$ &   \\\hline
$D^+\to \rho^0 \overline{K}^0\ell^+\nu_\ell$  &$39.54\pm5.02_{[\overline{K}_1(1270)^0]}$&$0.10\pm0.04_{[\overline{K}^*_2(1430)^0]}$&$34.20\pm4.36_{[\overline{K}_1(1270)^0]}$ &$0.07\pm0.03_{[\overline{K}^*_2(1430)^0]} $ \\
& $0.08\pm0.08_{[\overline{K}_1(1400)^0]}$  &  & $0.06\pm0.06_{[\overline{K}_1(1400)^0]}$ &   \\\hline
$D^+\to \omega \overline{K}^0\ell^+\nu_\ell$  &$21.65\pm2.78_{[\overline{K}_1(1270)^0]}$&$0.11\pm0.06_{[\overline{K}^*_2(1430)^0]}$&$18.72\pm2.41_{[\overline{K}_1(1270)^0]}$ &$0.07\pm0.04_{[\overline{K}^*_2(1430)^0]} $ \\
& $0.08\pm0.08_{[\overline{K}_1(1400)^0]}$  &  & $0.06\pm0.06_{[\overline{K}_1(1400)^0]}$ &   \\\hline
$D^+\to \overline{K}^{*0}\eta \ell^+\nu_\ell$  &$1.03\pm0.44_{[\overline{K}_1(1270)^0]}$&$0.02\pm0.01_{[\overline{K}^*_2(1430)^0]}$&$0.89\pm0.38_{[\overline{K}_1(1270)^0]}$ &$0.01\pm0.005_{[\overline{K}^*_2(1430)^0]} $ \\
& $0.09\pm0.09_{[\overline{K}_1(1400)^0]}$  &  & $0.07\pm0.07_{[\overline{K}_1(1400)^0]}$ &   \\\hline
$D^+\to \phi \overline{K}^0\ell^+\nu_\ell$  &$0.18\pm0.18_{[\overline{K}_1(1400)^0]}$&$0.001\pm0.0006_{[\overline{K}^*_2(1430)^0]}$&$0.14\pm0.14_{[\overline{K}_1(1400)^0]}$ &$0.0009\pm0.0004_{[\overline{K}^*_2(1430)^0]} $  \\\hline
$D^+\to \rho^+\pi^-\ell^+\nu_\ell$  &$2.41\pm1.39_{[a_1(1260)^0]}$&$0.16\pm0.06_{[a_2(1320)^0]}$&$2.12\pm1.24_{[a_1(1260)^0]}$ &$0.13\pm0.05_{[a_2(1320)^0]} $ \\
& $1.40\pm1.05_{[h_1(1170)^0]}$   &  & $1.25\pm0.94_{[[h_1(1170)^0]}$ &\\
&$0.03\pm0.03_{[h_1(1415)^0]}$ & &$0.02\pm0.02_{[[h_1(1415)^0]}$ &  \\\hline
$D^+\to \rho^-\pi^+\ell^+\nu_\ell$&$2.41\pm1.39_{[a_1(1260)^0]}$&$0.16\pm0.06_{[a_2(1320)^0]}$&$2.12\pm1.24_{[a_1(1260)^0]}$ &$0.13\pm0.05_{[a_2(1320)^0]} $ \\
& $1.40\pm1.05_{[h_1(1170)^0]}$   &  & $1.25\pm0.94_{[[h_1(1170)^0]}$ &\\
&$0.03\pm0.03_{[h_1(1415)^0]}$ & &$0.02\pm0.02_{[[h_1(1415)^0]}$ &  \\\hline
$D^+\to \omega\pi^0\ell^+\nu_\ell$  &$2.47\pm1.66_{[b_1(1235)^0]}$ & &$2.17\pm1.45_{[b_1(1235)^0]}$&   \\\hline
$D^+\to \rho^0\eta\ell^+\nu_\ell$  &$0.20\pm0.13_{[b_1(1235)^0]}$ & &$0.17\pm0.12_{[b_1(1235)^0]}$ &  \\\hline
$D^+\to \rho^0\pi^0\ell^+\nu_\ell$  &$1.40\pm1.05_{[h_1(1170)^0]}$ & &$1.26\pm0.95_{[h_1(1170)^0]}$&   \\
&$0.03\pm0.03_{[h_1(1415)^0]}$ & &$0.02\pm0.02_{[h_1(1415)^0]}$&  \\\hline
$D^+\to \omega\eta\ell^+\nu_\ell$  &$0.12\pm0.10_{[h_1(1170)^0]}$ & &$0.11\pm0.09_{[h_1(1170)^0]}$&   \\
&$0.007\pm0.007_{[h_1(1415)^0]}$ & &$0.006\pm0.006_{[h_1(1415)^0]}$&  \\\hline
$D^+\to \phi\eta\ell^+\nu_\ell$  &$0.004\pm0.004_{[h_1(1170)^0]}$ & &$0.004\pm0.004_{[h_1(1170)^0]}$&   \\
&$0.001\pm0.001_{[h_1(1415)^0]}$ & &$0.0008\pm0.0008_{[h_1(1415)^0]}$&  \\\hline
$D^+\to K^{*+}K^-\ell^+\nu_\ell$  &$0.10\pm0.05_{[a_1(1260)^0]}$&$0.0003\pm0.0001_{[a_2(1320)^0]}$&$0.09\pm0.04_{[a_1(1260)^0]}$ &$0.0002\pm0.00008_{[a_2(1320)^0]}$  \\
&$0.02\pm0.01_{[b_1(1235)^0]}$&$0.0006\pm0.0003_{[f_2(1270)]}$&$0.01\pm0.01_{[b_1(1235)^0]}$&$0.0005\pm0.0002_{[f_2(1270)]}$ \\
&$0.01\pm0.01_{[f_1(1285)]}$&$0.00003\pm0.00002_{[f_2'(1525)]}$&$0.009\pm0.009_{[f_1(1285)]}$&$0.00002\pm0.00001_{[f_2'(1525)]}$ \\
&$0.15\pm0.15_{[f_1(1420)]}$ & &$0.11\pm0.11_{[f_1(1420)]}$ & \\
&$0.05\pm0.05_{[h_1(1170)]}$ & &$0.05\pm0.04_{[h_1(1170)]}$&  \\
&$0.007\pm0.007_{[h_1(1415)]}$ & &$0.006\pm0.006_{[h_1(1415)]}$&  \\\hline
$D^+\to K^{*-}K^+\ell^+\nu_\ell$  &$0.10\pm0.05_{[a_1(1260)^0]}$&$0.0003\pm0.0001_{[a_2(1320)^0]}$&$0.09\pm0.04_{[a_1(1260)^0]}$ &$0.0002\pm0.00008_{[a_2(1320)^0]}$  \\
&$0.02\pm0.01_{[b_1(1235)^0]}$&$0.0006\pm0.0003_{[f_2(1270)]}$&$0.01\pm0.01_{[b_1(1235)^0]}$&$0.0005\pm0.0002_{[f_2(1270)]}$ \\
&$0.01\pm0.01_{[f_1(1285)]}$&$0.00003\pm0.00002_{[f_2'(1525)]}$&$0.009\pm0.009_{[f_1(1285)]}$&$0.00002\pm0.00001_{[f_2'(1525)]}$ \\
&$0.15\pm0.15_{[f_1(1420)]}$ & &$0.11\pm0.11_{[f_1(1420)]}$ & \\
&$0.05\pm0.05_{[h_1(1170)]}$ & &$0.05\pm0.04_{[h_1(1170)]}$&  \\
&$0.007\pm0.007_{[h_1(1415)]}$ & &$0.006\pm0.006_{[h_1(1415)]}$&  \\\hline
$D^+\to K^{*0}\overline{K}^0\ell^+\nu_\ell$&$0.09\pm0.04_{[a_1(1260)^0]}$&$0.0002\pm0.00009_{[a_2(1320)^0]}$&$0.08\pm0.04_{[a_1(1260)^0]}$ &$0.0002\pm0.00006_{[a_2(1320)^0]}$  \\
&$0.01\pm0.01_{[b_1(1235)^0]}$&$0.0005\pm0.0002_{[f_2(1270)]}$&$0.01\pm0.009_{[b_1(1235)^0]}$&$0.0004\pm0.0002_{[f_2(1270)]}$ \\
&$0.002\pm0.002_{[f_1(1285)]}$&$0.00003\pm0.00002_{[f_2'(1525)]}$&$0.002\pm0.002_{[f_1(1285)]}$&$0.00002\pm0.00001_{[f_2'(1525)]}$ \\
&$0.13\pm0.13_{[f_1(1420)]}$ & &$0.10\pm0.10_{[f_1(1420)]}$ & \\
&$0.05\pm0.05_{[h_1(1170)]}$ & &$0.04\pm0.04_{[h_1(1170)]}$&  \\
&$0.006\pm0.006_{[h_1(1415)]}$ & &$0.005\pm0.005_{[h_1(1415)]}$&  \\\hline
$D^+\to \overline{K}^{*0}K^0\ell^+\nu_\ell$  &$0.09\pm0.04_{[a_1(1260)^0]}$&$0.0002\pm0.00009_{[a_2(1320)^0]}$&$0.08\pm0.04_{[a_1(1260)^0]}$ &$0.0002\pm0.00006_{[a_2(1320)^0]}$  \\
&$0.01\pm0.01_{[b_1(1235)^0]}$&$0.0005\pm0.0002_{[f_2(1270)]}$&$0.01\pm0.009_{[b_1(1235)^0]}$&$0.0004\pm0.0002_{[f_2(1270)]}$ \\
&$0.002\pm0.002_{[f_1(1285)]}$ &$0.00003\pm0.00002_{[f_2'(1525)]}$&$0.002\pm0.002_{[f_1(1285)]}$&$0.00002\pm0.00001_{[f_2'(1525)]}$ \\
&$0.13\pm0.13_{[f_1(1420)]}$ & &$0.10\pm0.10_{[f_1(1420)]}$ & \\
&$0.05\pm0.05_{[h_1(1170)]}$ & &$0.04\pm0.04_{[h_1(1170)]}$&  \\
&$0.006\pm0.006_{[h_1(1415)]}$ & &$0.005\pm0.005_{[h_1(1415)]}$&  \\\hline
\end{tabular}}\label{Tab:D2PVlvAT2}
\end{table}

\begin{table}[h]
\renewcommand\arraystretch{1.15}
\tabcolsep 0.05in
\centering
\caption{SU(3) Branching ratios of  $D_s^+\to VP\ell^+\nu_{\ell}(\ell=e/\mu )$ decays within 1$\sigma$ error. The unit is $10^{-5}$ for the branching ratios. $\mathcal{B}_{[A]}$ and $\mathcal{B}_{[T]}$ denote the axial meson resonants and tensor meson resonants respectively. }\vspace{0.10cm}
{\footnotesize
\begin{tabular}{l|cc|cc}  \hline\hline
Decay modes & $\mathcal{B}_{[A]}(\ell=e)$  & $\mathcal{B}_{[T]} (\ell=e)$ & $\mathcal{B}_{[A]}(\ell=\mu)$ & $\mathcal{B}_{[T]}(\ell=\mu)$ \\\hline
$D_s^+\to K^{*-}\pi^+\ell^+\nu_\ell$  &$1.82\pm0.87_{[K_1(1270)^0]}$&$0.06\pm0.02_{[K^*_2(1430)^0]}$ &$1.64\pm0.78_{[K_1(1270)^0]}$    &$0.04\pm0.02_{[K^*_2(1430)^0]} $ \\
& $0.19\pm0.19_{[K_1(1400)^0]}$  & & $0.16\pm0.16_{[K_1(1400)^0]}$ &   \\\hline
$D_s^+\to \overline{K}^{*0}\pi^0\ell^+\nu_\ell$  &$0.91\pm0.43_{[K_1(1270)^0]}$&$0.03\pm0.01_{[K^*_2(1430)^0]}$ &$0.81\pm0.39_{[K_1(1270)^0]}$    &$0.02\pm0.008_{[K^*_2(1430)^0]} $ \\
& $0.09\pm0.09_{[K_1(1400)^0]}$  & & $0.08\pm0.08_{[K_1(1400)^0]}$ &   \\\hline
$D_s^+\to \rho^+ K^-\ell^+\nu_\ell$  &$4.00\pm0.54_{[K_1(1270)^0]}$&$0.02\pm0.009_{[K^*_2(1430)^0]}$ &$3.59\pm0.48_{[K_1(1270)^0]}$    &$0.02\pm0.007_{[K^*_2(1430)^0]} $ \\
& $0.01\pm0.01_{[K_1(1400)^0]}$  & & $0.008\pm0.008_{[K_1(1400)^0]}$ &   \\\hline
$D_s^+\to \rho^0 \overline{K}^0\ell^+\nu_\ell$  &$1.92\pm0.25_{[K_1(1270)^0]}$&$0.01\pm0.004_{[K^*_2(1430)^0]}$ &$1.72\pm0.23_{[K_1(1270)^0]}$    &$0.008\pm0.003_{[K^*_2(1430)^0]} $ \\
& $0.005\pm0.005_{[K_1(1400)^0]}$  & & $0.004\pm0.004_{[K_1(1400)^0]}$ &   \\\hline
$D_s^+\to \omega \overline{K}^0\ell^+\nu_\ell$  &$1.04\pm0.14_{[K_1(1270)^0]}$&$0.01\pm0.006_{[K^*_2(1430)^0]}$ &$0.94\pm0.12_{[K_1(1270)^0]}$    &$0.008\pm0.004_{[K^*_2(1430)^0]} $ \\
& $0.005\pm0.005_{[K_1(1400)^0]}$  & & $0.004\pm0.004_{[K_1(1400)^0]}$ &   \\\hline
$D_s^+\to \overline{K}^{*0}\eta \ell^+\nu_\ell$ &$0.05\pm0.02_{[K_1(1270)^0]}$&$0.002\pm0.0009_{[K^*_2(1430)^0]}$ &$0.05\pm0.02_{[K_1(1270)^0]}$    &$0.001\pm0.0006_{[K^*_2(1430)^0]} $ \\
& $0.005\pm0.005_{[K_1(1400)^0]}$  & & $0.004\pm0.004_{[K_1(1400)^0]}$ &   \\\hline
$D_s^+\to \phi \overline{K}^0\ell^+\nu_\ell$  &$0.01\pm0.01_{[K_1(1400)^0]}$&$0.0001\pm0.00006_{[K^*_2(1430)^0]}$&$0.009\pm0.009_{[K_1(1400)^0]}$ &$0.0001\pm0.00005_{[K^*_2(1430)^0]} $  \\\hline
$D_s^+\to \rho^+\pi^-\ell^+\nu_\ell$  &$4.21\pm4.21_{[h_1(1170)]}$ & &$3.86\pm3.86_{[h_1(1170)]}$&     \\
&$3.47\pm3.47_{[h_1(1415)]}$ & &$2.92\pm2.92_{[[h_1(1415)]}$ &\\\hline
$D_s^+\to \rho^-\pi^+\ell^+\nu_\ell$&$4.21\pm4.21_{[h_1(1170)]}$ & &$3.86\pm3.86_{[h_1(1170)]}$&     \\
&$3.47\pm3.47_{[h_1(1415)]}$ & &$2.92\pm2.92_{[[h_1(1415)]}$ &\\\hline
$D_s^+\to \rho^0\pi^0\ell^+\nu_\ell$  &$4.23\pm4.23_{[h_1(1170)]}$ & &$3.89\pm3.88_{[h_1(1170)]}$&     \\
&$3.48\pm3.48_{[h_1(1415)]}$ & &$2.93\pm2.93_{[[h_1(1415)]}$ &\\\hline
$D_s^+\to \omega\eta\ell^+\nu_\ell$  &$0.41\pm0.41_{[h_1(1170)]}$ & &$0.37\pm0.37_{[h_1(1170)]}$&     \\
&$0.89\pm0.89_{[h_1(1415)]}$ & &$0.75\pm0.75_{[[h_1(1415)]}$ &\\\hline
$D_s^+\to K^{*+}K^-\ell^+\nu_\ell$  &$0.51\pm0.46_{[f_1(1285)]}$&$0.0007\pm0.0004_{[f_2(1270)]}$&$0.46\pm0.41_{[f_1(1285)]}$ &$0.0006\pm0.0003_{[f_2(1270)]}$  \\
&$5.24\pm5.24_{[f_1(1420)]}$&$0.11\pm0.04_{[f_2'(1525)]}$&$4.36\pm4.36_{[f_1(1420)]}$ &$0.07\pm0.03_{[f_2'(1525)]}$ \\
&$0.37\pm0.37_{[h_1(1170)]}$ & &$0.34\pm0.34_{[h_1(1170)]}$ & \\
&$4.04\pm3.70_{[h_1(1415)]}$ & &$3.39\pm3.11_{[h_1(1415)]}$&  \\\hline
$D_s^+\to K^{*-}K^+\ell^+\nu_\ell$&$0.51\pm0.46_{[f_1(1285)]}$&$0.0007\pm0.0004_{[f_2(1270)]}$&$0.46\pm0.41_{[f_1(1285)]}$ &$0.0006\pm0.0003_{[f_2(1270)]}$  \\
&$5.24\pm5.24_{[f_1(1420)]}$&$0.11\pm0.04_{[f_2'(1525)]}$&$4.36\pm4.36_{[f_1(1420)]}$ &$0.07\pm0.03_{[f_2'(1525)]}$ \\
&$0.37\pm0.37_{[h_1(1170)]}$ & &$0.34\pm0.34_{[h_1(1170)]}$ & \\
&$4.04\pm3.70_{[h_1(1415)]}$ & &$3.39\pm3.11_{[h_1(1415)]}$&  \\\hline
$D_s^+\to K^{*0}\overline{K}^0\ell^+\nu_\ell$  &$0.11\pm0.10_{[f_1(1285)]}$&$0.0007\pm0.0003_{[f_2(1270)]}$&$0.10\pm0.09_{[f_1(1285)]}$ &$0.0005\pm0.0003_{[f_2(1270)]}$  \\
&$4.71\pm4.71_{[f_1(1420)]}$&$0.10\pm0.04_{[f_2'(1525)]}$&$3.92\pm3.92_{[f_1(1420)]}$ &$0.07\pm0.02_{[f_2'(1525)]}$ \\
&$0.35\pm0.35_{[h_1(1170)]}$ & &$0.32\pm0.32_{[h_1(1170)]}$ & \\
&$3.43\pm3.16_{[h_1(1415)]}$ & &$2.88\pm2.65_{[h_1(1415)]}$&  \\\hline
$D_s^+\to \overline{K}^{*0}K^0\ell^+\nu_\ell$  &$0.11\pm0.10_{[f_1(1285)]}$&$0.0007\pm0.0003_{[f_2(1270)]}$&$0.10\pm.09_{[f_1(1285)]}$ &$0.0005\pm0.0003_{[f_2(1270)]}$  \\
&$4.71\pm4.71_{[f_1(1420)]}$&$0.10\pm0.04_{[f_2'(1525)]}$&$3.92\pm3.92_{[f_1(1420)]}$ &$0.07\pm0.02_{[f_2'(1525)]}$ \\
&$0.35\pm0.35_{[h_1(1170)]}$ & &$0.32\pm0.32_{[h_1(1170)]}$ & \\
&$3.43\pm3.16_{[h_1(1415)]}$ & &$2.88\pm2.65_{[h_1(1415)]}$&  \\\hline
\end{tabular}}\label{Tab:D2PVlvAT3}
\end{table}

Comparing the experimental upper limits in Eqs. (\ref{Eq:BrB2PVlvA1}-\ref{Eq:BrB2PVlvA2}) and our predictions in Tab. \ref{Tab:D2PVlvAT2}, we can see our predictions of  $\mathcal{B}(D^+\to b_1(1235)^0e^+\nu_e,~b_1(1235)^0\to \omega\pi^0)$ and $\mathcal{B}(D^0\to b_1(1235)^-e^+\nu_e,~b_1(1235)^-\to \omega\pi^-)$ are much smaller than the experimental  upper limits.
After we completed the analysis of the axial-vector resonant contributions,  we saw the experimental report regarding $\mathcal{B}(D^+\to b_1(1235)^0e^+\nu_e,~b_1(1235)^0\to \omega\pi^0)$ and $\mathcal{B}(D^0\to b_1(1235)^-e^+\nu_e,~b_1(1235)^-\to \omega\pi^-)$ from BESIII \cite{BESIII:2024pwp}
\begin{eqnarray}
\mathcal{B}(D^+\to b_1(1235)^0e^+\nu_e,~b_1(1235)^0\to \omega\pi^0)&=&(1.16\pm0.44\pm0.16)\times10^{-4},\\
\mathcal{B}(D^0\to b_1(1235)^-e^+\nu_e,~b_1(1235)^-\to \omega\pi^-)&=&(0.72\pm0.18^{+0.06}_{-0.08})\times10^{-4}.
\end{eqnarray}
The above data are slightly larger than our predictions in Tab. \ref{Tab:D2PVlvAT2}, and they are consistent within $2\sigma$ errors.
As mentioned earlier, for some branching ratio predictions, due to poor experimental data at present and our conservative choice of the non-perturbative parameters, their errors are as large as their central values, which means that we can only give the upper limits.
Many predictions of $\mathcal{B}(D\to A\ell^+\nu_\ell,A\to VP)$ are on the order of $\mathcal{O}(10^{-5}-10^{-4})$, and they are expected to be measured in the near future.


\section{Semileptonic $D\rightarrow T(T\to PV)\ell^+\nu_\ell $ decays} \label{Sec:D2PVlvT}
\subsection{Semileptonic $D\rightarrow T\ell^+\nu_\ell $ decays} \label{Sec:D2Tlv}

\subsubsection{Theoretical framework for the $D\to T \ell^+\nu_\ell$ decays}
With the same  effective Hamiltonian given in Eq. (\ref{Heff}),
the decay amplitudes of the $D\to T\ell^+\nu_\ell$ decays  are similar to those of the $D\to A\ell^+\nu_\ell$ decays given in Eqs. (\ref{Eq:MD2Alv1}-\ref{Eq:MD2Alv3}) by replacing $A$ with $T$.

The form factors of the  $D\to T$ transitions are similar to those of the  $B\to T$ transitions  \cite{Chen:2021ywv}
\begin{eqnarray}
\left<T(p,\varepsilon^{*})\left|\bar{u}_k\gamma_{\mu}(1-\gamma_5)b\right|D(p_D)\right>
&=&\frac{2iV^T(q^2)}{m_D+m_T}\epsilon_{\mu\nu\alpha\beta}e^{*\nu}p^\alpha_Dp^\beta\nonumber\\
&&+2m_T\frac{e^*\cdot q}{q^2}q_\mu A^T_0(q^2)+(m_D+m_T)\Big(e^*_\mu-\frac{e^*\cdot q}{q^2}q_\mu\Big)A^T_1(q^2)\nonumber\\
&&-\frac{e^*\cdot q}{m_D+m_T}\Big((p_D+p)_\mu-\frac{m_D^2-m^2_T}{q^2}q_\mu\Big)A^T_2(q^2),
\end{eqnarray}
where $q^2\equiv(p_D-p)^2$, $V^T(q^2)$, and $A^T_{0,1,2}(q^2)$ are the form factors of the $D\to T$ transition,  $m_{T}$ is the mass of  the tensor mesons, and $e^{*\nu}\equiv \frac{\varepsilon^{*\mu\nu}\cdot p_{D\mu}}{m_D}$. The hadronic amplitudes of $D\to T\ell^+\nu_{\ell}$ decays can then  be written as
\begin{eqnarray}
H^T_{\pm} &=&\frac{2|\vec{p}_T|}{\sqrt{6}m_T}\left[(m_{D_q}+m_T)A^T_1(q^2)\mp\frac{2m_{D_q}|\vec{p}_T|}{(m_{D_q}+m_T)}V^T(q^2)\right], \\
H^T_{0}&=& \frac{|\vec{p}_T|}{\sqrt{2}m_T}\frac{1}{2m_T\sqrt{q^2}}\left[(m_{D_q}^2-m_T^2-q^2)(m_{D_q}+m_T)A^T_1(q^2)-\frac{4m_{D_q}^2|\vec{p}_T|^2}{m_{D_q}+m_T}A^T_2(q^2)\right], \\
H^T_{t}&=& \frac{|\vec{p}_T|}{\sqrt{2}m_T}\frac{2m_{D_q}|\vec{p}_T|}{\sqrt{q^2}}A^T_0(q^2),
\end{eqnarray}
where $|\vec{p}_T|\equiv\sqrt{\lambda(m_{D_q}^2,m_T^2,q^2)}/2m_{D_q}$ with $\lambda(a,b,c)=a^2+b^2+c^2-2ab-2ac-2bc$.

Similar to the $D\to A\ell^+\nu_{\ell}$ decays, the hadronic amplitude relations or form factor relations of the $D\to T\ell^+\nu_{\ell}$ decays can be obtained from Eq. (\ref{Eq:AmpD2AlvSU3}) with $M=T$ by using the SU(3) flavor symmetry approach, and are listed in Tab. \ref{D2TlvAmp}. The SU(3) amplitudes of the $D\to T\ell^+\nu_{\ell}$ decays contain non-perturbation coefficients $T_{1,2,3,4}$; if  neglecting the SU(3) flavor breaking $c^T_1$ and $c^T_2$  terms, there is only one non-perturbation parameter $T_1=T_2=T_3=T_4=c^T_0$.

To date,  no experimental data have been available for the $D\to T\ell^+\nu_\ell$ decays, so we can not extract the non-perturbative coefficients from the experimental data.
In Ref. \cite{Chen:2021ywv}, some form factors of $D\to T$  were calculated in the light-front quark model.
Comparing the form factors among $D\to a_2$, $D\to K^*_2$, and$D_s\to K^*_2$ transitions in Tab. 3 of Ref. \cite{Chen:2021ywv},  no significant difference is observed. Therefore, we take $F(0)$ and $b_1$ of the $D\to a_2$ transition as input parameters, and the form factors of the other decays can be obtained by the relations in Tab. \ref{D2TlvAmp} after ignoring the SU(3) flavor breaking effects.
Our branching ratio predictions for the $D\to T\ell^+\nu_\ell$ decays are listed in Tab. \ref{Tab:BrD2Tlv}.

One can see that
the predictions of  $\mathcal{B}(D^0\to K^*_2(1430)^-\ell^+\nu_\ell)$, $\mathcal{B}(D^+\to \overline{K}^*_2(1430)^0\ell^+\nu_\ell)$, $\mathcal{B}(D_s^+\to f_2(1270)\ell^+\nu_\ell)$, $\mathcal{B}(D_s^+\to f_2'(1525)\ell^+\nu_\ell)$, and $\mathcal{B}(D^+\to f_2(1270)e^+\nu_e)$
 reached the order of $10^{-5}$. Other branching ratios are obviously suppressed by the CKM matrix element $V^*_{cd}$.

\begin{table}[h]
\renewcommand\arraystretch{1.2}
\tabcolsep 0.15in
\centering
\caption{The hadronic amplitudes for  $D\to T\ell^+\nu$ decays. $T_1\equiv c^T_0+c^T_1-2c^T_2$, $T_2\equiv c^T_0-2c^T_1-2c^T_2$, $T_3\equiv c^T_0+c^T_1+c^T_2$, $T_4\equiv c^T_0-2c^T_1+c^T_2$. $T_1=T_2=T_3=T_4=c^T_0$ if  neglecting the SU(3) flavor breaking $c^T_1$ and $c^T_2$  terms.
}\vspace{0.1cm}
{\footnotesize
\begin{tabular}{lc|lc}  \hline\hline
Hadronic helicity amplitudes  & SU(3) amplitudes & Hadronic helicity amplitudes  & SU(3) amplitudes \\\hline
$H(D^0\to K^*_2(1430)^-\ell^+\nu_\ell)$             &$T_1~V^*_{cs}$                    &  $H(D^0\to a_2(1320)^-\ell^+\nu_\ell)$            &$T_3~V^*_{cd}$                                      \\
$H(D^+\to \overline{K}^*_2(1430)^0\ell^+\nu_\ell)$  &$T_1~V^*_{cs}$                    &  $H(D^+\to a_2(1320)^0\ell^+\nu_\ell)$            &$-\frac{1}{\sqrt{2}}T_3~V^*_{cd}$                      \\
$H(D_s^+\to f_2(1270)\ell^+\nu_\ell)$               &$T_2~V^*_{cs}~sin\theta_{f_2}$     &  $H(D^+\to f_2(1270)\ell^+\nu_\ell)$             &$\frac{1}{\sqrt{2}}T_3~V^*_{cd}~cos\theta_{f_2}$        \\
$H(D_s^+\to f_2'(1525)\ell^+\nu_\ell)$              &$-T_2~V^*_{cs}~cos\theta_{f_2}$    &  $H(D^+\to f_2'(1525)\ell^+\nu_\ell)$            &$\frac{1}{\sqrt{2}}T_3~V^*_{cd}~sin\theta_{f_2}$         \\
                                                    &                                 &  $H(D_s^+\to K^*_2(1430)^0\ell^+\nu_\ell)$         &$T_4~V^*_{cd}$                                            \\ \hline
\end{tabular}} \label{D2TlvAmp}
%
\renewcommand\arraystretch{1.2}
\tabcolsep 0.25in
\centering
\caption{Branching ratio predictions  of the  $D\to T\ell^+\nu_{\ell}$ decays within 1$\sigma$ error (in units of $10^{-5}$). }\vspace{0.1cm}
{\footnotesize
\begin{tabular}{lccc}  \hline\hline
Branching ratios                                                 &  Predictions $(\ell=e)$          & Predictions $(\ell=\mu)$\\\hline
$\mathcal{B}(D^0\to K^*_2(1430)^-\ell^+\nu_\ell)$                &$1.36\pm0.49$                     &$0.91\pm0.31$                  \\
$\mathcal{B}(D^+\to \overline{K}^*_2(1430)^0\ell^+\nu_\ell)$     &$3.40\pm1.21$                     &$2.25\pm0.76$                   \\
$\mathcal{B}(D_s^+\to f_2(1270)\ell^+\nu_\ell)$                  &$1.04\pm0.54$                     &$0.87\pm0.44$                    \\
$\mathcal{B}(D_s^+\to f_2'(1525)\ell^+\nu_\ell)$                 &$1.83\pm0.69$                     &$1.25\pm0.45$                    \\
$\mathcal{B}(D^0\to a_2(1320)^-\ell^+\nu_\ell)$                  &$0.35\pm0.13$                     &$0.27\pm0.09$                  \\
$\mathcal{B}(D^+\to a_2(1320)^0\ell^+\nu_\ell)$                  &$0.47\pm0.17$                     &$0.36\pm0.12$                   \\
$\mathcal{B}(D^+\to f_2(1270)\ell^+\nu_\ell)$                    &$0.78\pm0.29$                     &$0.62\pm0.21$                    \\
$\mathcal{B}(D^+\to f_2'(1525)\ell^+\nu_\ell)$                   &$(5.35\pm2.83)\times10^{-4}$      &$(2.90\pm1.48)\times10^{-4}$     \\
$\mathcal{B}(D_s^+\to K^*_2(1430)^0\ell^+\nu_\ell)$              &$0.34\pm0.13$                     &$0.26\pm0.09$                    \\ \hline
\end{tabular}\label{Tab:BrD2Tlv}}
\end{table}

\subsection{Nonleptonic $T \to VP$ decays}

The branching ratios of the $T\to VP$ decays can be written as  \cite{Li:2015xka}
\begin{equation}
\mathcal{B}(T\to VP)=\frac{\tau_T|\lambda_{TVP}|^2(q'_T)^5}{20\pi},
\end{equation}
where $q'_T\equiv\frac{1}{2m_T}\lambda^{1/2}(m_T^2,m_P^2,m_V^2)$, and  $\lambda_{TVP}$ is the coefficient of the $TVP$ vertex, and $\tau_T$ is the lifetime of the tensor meson.
Since most of the $T\to VP$  modes  decay by small phases,  the mass distribution of tensor resonances and vector mesons is considered in all $T\to VP$ decays, which are similar to the $A\to VP$ decays in Eq. (\ref{Eq:GammaA2VP2}) by replacing $A$ with $T$.

Under the SU(3) flavor symmetry, the $\lambda_{TVP}$  of different $T\to VP$ decays can be related.
The $TVP$ vertex coefficients  can be obtained by replacing $A$ with $T$ in Eq. (\ref{Eq:LambdaAVP}), and the results  are listed in Tab. \ref{Tab:T2PVAmp}.
After ignoring the SU(3) flavor breaking terms, we see that all the vertex coefficients of the $T\to VP$ decays are related to the non-perturbative coefficients $c'^T_0$. Apart from the mixing angle $\theta_P$, there remains only  $\theta_{f_2}$ in the $f_2(1270)/f_2'(1525)\to VP$ decay modes.
\begin{table}[b]
\renewcommand\arraystretch{1.0}
\tabcolsep 0.1in
\centering
\caption{The vertex coefficients of $T\to VP$ decays under the SU(3) flavor symmetry. $a^T_1=c'^T_0-c'^T_1+c'^T_3$, $a^T_2=c'^T_0-c'^T_1-2c'^T_3$, $a^T_3=c'^T_0+2c'^T_1+c'^T_3$, $a^T_4=c'^T_0+2c'^T_1-2c'^T_3$, $a^T_5=c'^T_0-4c'^T_1+c'^T_3$.
$a^T_1=a^T_2=a^T_3=a^T_4=a^T_5=c'^T_0$  if neglecting the SU(3) flavor breaking $c'^T_1$ and $c'^T_3$ terms.}\vspace{0.08cm}
{\footnotesize
\begin{tabular}{lc|lc}  \hline\hline
Helicity amplitudes  & SU(3) flavor amplitudes &Helicity amplitudes  & SU(3) flavor amplitudes\\\hline
$\overline{K}_2^*(1430)^0\to \rho^+ K^-$              &$-a^T_1$                                     &  $\overline{K}_2^*(1430)^0\to \overline{K}^{*0}\eta$           &$\frac{1}{\sqrt{6}}\cos\theta_P(a^T_1+2a^T_2)-\frac{1}{\sqrt{3}}\sin\theta_P(a^T_1-a^T_2)$              \\
$\overline{K}_2^*(1430)^0\to \rho^0 \overline{K}^0$   &$\frac{1}{\sqrt{2}}a^T_1$                    &  $K_2^*(1430)^-\to K^{*-}\eta$                                &    $\frac{1}{\sqrt{6}}\cos\theta_P(a^T_1+2a^T_2)-\frac{1}{\sqrt{3}}\sin\theta_P(a^T_1-a^T_2)$        \\
$K_2^*(1430)^-\to \rho^0 K^-$                         &$-\frac{1}{\sqrt{2}}a^T_1$                   &  $\overline{K}_2^*(1430)^0\to \overline{K}^{*0}\eta'$          &$\frac{1}{\sqrt{6}}\sin\theta_P(a^T_1+2a^T_2)+\frac{1}{\sqrt{3}}\cos\theta_P(a^T_1-a^T_2)$                  \\
$K_2^*(1430)^-\to \rho^- \overline{K}^0$              &$-a^T_1$                                     &  $K_2^*(1430)^-\to  K^{*-}\eta'$                              & $\frac{1}{\sqrt{6}}\sin\theta_P(a^T_1+2a^T_2)+\frac{1}{\sqrt{3}}\cos\theta_P(a^T_1-a^T_2)$   \\
$\overline{K}_2^*(1430)^0\to K^{*-}\pi^+$             &$a^T_1$                                      &  $\overline{K}_2^*(1430)^0\to \omega \overline{K}^0$                      &$-\frac{1}{\sqrt{2}}a^T_1$                      \\

$\overline{K}_2^*(1430)^0\to \overline{K}^{*0}\pi^0$  &$-\frac{1}{\sqrt{2}}a^T_1$                   &   $K_2^*(1430)^-\to \omega K^-$                         &    $-\frac{1}{\sqrt{2}}a^T_1$                      \\
$K_2^*(1430)^-\to K^{*-}\pi^0$                       &$\frac{1}{\sqrt{2}}a^T_1$                     &    $\overline{K}_2^*(1430)^0\to \phi \overline{K}^0$                        &$a^T_2$                                            \\
$K_2^*(1430)^-\to \overline{K}^{*0}\pi^-$            &$a^T_1$                                       &    $K_2^*(1430)^-\to\phi K^-$                           &    $a^T_2$                                                \\\hline
$a_2(1320)^0\to \rho^+\pi^-$             & $\sqrt{2}a^T_3$                   &  $a_2(1320)^0\to K^{*0}\overline{K}^0$       &    $-\frac{1}{\sqrt{2}}a^T_4$       \\
$a_2(1320)^0\to \rho^-\pi^+$             & $-\sqrt{2}a^T_3$                  &  $a_2(1320)^0\to \overline{K}^{*0}K^{0}$     &    $\frac{1}{\sqrt{2}}a^T_4$        \\
$a_2(1320)^-\to \rho^-\pi^0$             & $\sqrt{2}a^T_3$                   &   $a_2(1320)^0\to K^{*+}K^{-}$               &    $\frac{1}{\sqrt{2}}a^T_4$         \\
$a_2(1320)^-\to \rho^0\pi^-$             & $-\sqrt{2}a^T_3$                  &  $a_2(1320)^0\to K^{*-}K^{+}$                &    $-\frac{1}{\sqrt{2}}a^T_4$        \\
                                            &                                 &  $a_2(1320)^-\to K^{*0}K^-$                  & $a^T_4$              \\

                                            &                                 &  $a_2(1320)^-\to K^{*-}K^0 $                 &$-a^T_4$              \\\hline

$f_2(1270) \to K^{*+}K^-$                  &    $a^T_4 \frac{\cos\theta_{f_2}}{\sqrt{2}}-a^T_5 \sin\theta_{f_2}$        & $f_2'(1525) \to K^{*+}K^-$               &    $a^T_4 \frac{\sin\theta_{f_2}}{\sqrt{2}}+a^T_5 \cos\theta_{f_2}$         \\
$f_2(1270) \to K^{*-}K^+$                  &    $-a^T_4 \frac{\cos\theta_{f_2}}{\sqrt{2}}+a^T_5 \sin\theta_{f_2}$       & $f_2'(1525) \to K^{*-}K^+$               &    $-a^T_4 \frac{\sin\theta_{f_2}}{\sqrt{2}}-a^T_5\cos\theta_{f_2}$         \\
$f_2(1270) \to K^{*0}\overline{K}^0$       &    $a^T_4 \frac{\cos\theta_{f_2}}{\sqrt{2}}-a^T_5 \sin\theta_{f_2}$        & $f_2'(1525) \to K^{*0}\overline{K}^0$    &    $a^T_4 \frac{\sin\theta_{f_2}}{\sqrt{2}}+a^T_5 \cos\theta_{f_2}$          \\
$f_2(1270) \to \overline{K}^{*0}K^0$       &    $-a^T_4\frac{\cos\theta_{f_2}}{\sqrt{2}}+a^T_5 \sin\theta_{f_2}$       & $f_2'(1525) \to\overline{K}^{*0}K^0$     &    $-a^T_4\frac{\sin\theta_{f_2}}{\sqrt{2}}-a^T_5 \cos\theta_{f_2}$          \\\hline
\end{tabular}} \label{Tab:T2PVAmp}
\end{table}

Now only some $T\to VP$ decays have been measured \cite{PDG2022}
\begin{eqnarray}
\mathcal{B}(K_2^*(1430)\to K^*\pi )&=&(24.7\pm1.5)\%,\\
\mathcal{B}(K_2^*(1430)\to \rho K  )&=&(8.7\pm0.8)\%,\\
\mathcal{B}(K_2^*(1430)\to \omega K  )&=&(2.9\pm0.8)\%,\\
\mathcal{B}(a_2(1320)\to \rho \pi )&=&(70.1\pm2.7)\%.
\end{eqnarray}
The above experimental data could be used to determine the $TVP$ vertex coefficients $\lambda_{TVP}$, and we obtain the following:  $|\lambda_{TVP}|=8.22\pm0.33$ from  $K_2^*(1430)\to K^*\pi$, $|\lambda_{TVP}|=7.14\pm0.39$ from  $K_2^*(1430)\to  \rho K $, $|\lambda_{TVP}|=8.81\pm1.32$ from  $K_2^*(1430)\to \omega K$, and $|\lambda_{TVP}|=8.44\pm0.23$ from  $a_2(1320)\to \rho \pi$. One can see that although $|\lambda_{TVP}|$ from  $K_2^*(1430)\to  \rho K $ is slightly small, the other three values are  coincident with each other within $1\sigma$ error.

In our numerical results, we use each constrained $\lambda_{TVP}$ to predict their specific processes. For examples, $|\lambda_{TVP}|=8.22\pm0.33$ from  $K_2^*(1430)\to K^*\pi$ is used to predict the branching ratios of the $\overline{K}_2^*(1430)^0\to K^{*-}\pi^+$, $\overline{K}_2^*(1430)^0\to \overline{K}^{*0}\pi^0$, $K_2^*(1430)^-\to K^{*-}\pi^0$, $K_2^*(1430)^-\to \overline{K}^{*0}\pi^-$ decays, and $|\lambda_{TVP}|=8.44\pm0.23$ from  $a_2(1320)\to \rho \pi$ is used to predict the branching ratios of the $a_2(1320)^0\to \rho^+\pi^-$, $a_2(1320)^0\to \rho^-\pi^+$, $a_2(1320)^-\to \rho^-\pi^0$, $a_2(1320)^-\to \rho^0\pi^-$ decays.
For the as-yet-unmeasured processes, for examples, $K_2^*(1430)\to K^{*}\eta$, $a_2(1320)\to K^{*}K$ and $f_2(1270) \to K^{*}K$,   we will use $|\lambda_{TVP}|=8.22\pm0.33$ from  $K_2^*(1430)\to K^*\pi$ to predict their branching ratios.

Our predictions  are listed in Tab. \ref{Tab:NunmT2VP}.
One can see that $\mathcal{B}(K^*_2(1430)\to \rho K)$, $\mathcal{B}(K^*_2(1430)\to K^*\pi)$,
 $\mathcal{B}(K^*_2(1430)\to \omega K)$,
$\mathcal{B}(a_2(1320)\to \rho\pi)$, and  $\mathcal{B}(f'_2(1525)\to K^*K)$ are on the order of $10^{-2}-10^{-1}$; other processes are strongly suppressed by their small phase spaces.
\begin{table}[th]
\renewcommand\arraystretch{0.96}
\tabcolsep 0.15in
\centering
\caption{Branching ratios of the $T\to VP$ decays within $1\sigma$ error (in units of $10^{-2}$). $^E$denotes the experimental data. }\vspace{0.08cm}
\begin{tabular}{lc|lc}  \hline\hline
  & Our predictions &   & Our predictions\\\hline
$\mathcal{B}(\overline{K}_2^*(1430)^0\to \rho^+ K^-)$              &$6.07\pm0.63$                                                                       &  $\mathcal{B}(\overline{K}_2^*(1430)^0\to \overline{K}^{*0}\eta)$           &$0.56\pm0.08$              \\
$\mathcal{B}(\overline{K}_2^*(1430)^0\to \rho^0 \overline{K}^0)$   &$2.93\pm0.31$                                                                       &  $\mathcal{B}(K_2^*(1430)^-\to K^{*-}\eta)$                                &    $0.56\pm0.08$              \\
$\mathcal{B}(K_2^*(1430)^-\to \rho^0 K^-)$                         &$2.86\pm0.30$                                                                       &  $\mathcal{B}(\overline{K}_2^*(1430)\to K^*\eta)$          &$0.56\pm0.08$                \\
$\mathcal{B}(K_2^*(1430)^-\to \rho^- \overline{K}^0)$              &$5.57\pm0.58$                                                                       &  $\mathcal{B}(\overline{K}_2^*(1430)^0\to \omega \overline{K}^0)$          &$2.94\pm0.85$                      \\
$\mathcal{B}(K_2^*(1430)\to \rho K)$                    &$8.70\pm0.80^E$                                                    &   $\mathcal{B}(K_2^*(1430)^-\to \omega K^-)$                               & $2.88\pm0.84$                      \\
$\mathcal{B}(\overline{K}_2^*(1430)^0\to K^{*-}\pi^+)$             &$17.01\pm1.18$                                                                      &   $\mathcal{B}(K_2^*(1430)\to \omega K)$
                          & $2.90\pm0.80^E$                      \\

$\mathcal{B}(\overline{K}_2^*(1430)^0\to \overline{K}^{*0}\pi^0)$  &$8.34\pm0.58$                                                                       &   $\mathcal{B}(\overline{K}_2^*(1430)^0\to \phi \overline{K}^0)$           & $0.040\pm0.008$        \\
$\mathcal{B}(K_2^*(1430)^-\to K^{*-}\pi^0)$                       &$8.25\pm0.58$                                                                        &   $\mathcal{B}(K_2^*(1430)^-\to\phi K^-)$                                  & $0.038\pm0.008$          \\
$\mathcal{B}(K_2^*(1430)^-\to \overline{K}^{*0}\pi^-)$            &$15.82\pm1.10$                                                                       &   $\mathcal{B}(K_2^*(1430)\to\phi K)$                                  & $0.039\pm0.007$          \\
$\mathcal{B}(K_2^*(1430)\to K^{*}\pi)$            &$24.70\pm1.50^E$                                                          \\  \hline
$\mathcal{B}(a_2(1320)^0\to \rho^+\pi^-)$             & $34.99\pm1.37$                   &  $\mathcal{B}(a_2(1320)^0\to K^{*0}\overline{K}^0)$       &    $0.043\pm0.005$       \\
$\mathcal{B}(a_2(1320)^0\to \rho^-\pi^+)$             & $34.99\pm1.37$                  &  $\mathcal{B}(a_2(1320)^0\to \overline{K}^{*0}K^{0})$     &    $0.043\pm0.005$        \\
$\mathcal{B}(a_2(1320)^-\to \rho^-\pi^0)$             & $35.33\pm1.38$                   &   $\mathcal{B}(a_2(1320)^0\to K^{*+}K^{-})$               &    $0.055\pm0.006$         \\
$\mathcal{B}(a_2(1320)^-\to \rho^0\pi^-)$             & $34.99\pm1.40$                  &  $\mathcal{B}(a_2(1320)^0\to K^{*-}K^{+})$                &    $0.055\pm0.006$        \\
$\mathcal{B}(a_2(1320)\to \rho\pi)$             & $70.10\pm2.70^E$        &  $\mathcal{B}(a_2(1320)^-\to K^{*0}K^-)$                  & $0.094\pm0.010$              \\

                                            &                                 &  $\mathcal{B}(a_2(1320)^-\to K^{*-}K^0 )$                 &$0.10\pm0.01$              \\

                                            &                                 &  $\mathcal{B}(a_2(1320)\to K^{*}K )$                 &$0.19\pm0.02$              \\\hline

$\mathcal{B}(f_2(1270) \to K^{*+}K^-)$                  &    $0.072\pm0.013$        & $\mathcal{B}(f_2'(1525) \to K^{*+}K^-)$               &    $5.89\pm0.75$         \\
$\mathcal{B}(f_2(1270) \to K^{*-}K^+)$                  &    $0.072\pm0.013$       & $\mathcal{B}(f_2'(1525) \to K^{*-}K^+)$               &    $5.89\pm0.75$         \\
$\mathcal{B}(f_2(1270) \to K^{*0}\overline{K}^0)$       &    $0.064\pm0.011$        & $\mathcal{B}(f_2'(1525) \to K^{*0}\overline{K}^0)$    &    $5.25\pm0.66$          \\
$\mathcal{B}(f_2(1270) \to \overline{K}^{*0}K^0)$       &    $0.064\pm0.011$       & $\mathcal{B}(f_2'(1525) \to\overline{K}^{*0}K^0)$     &    $5.25\pm0.66$          \\
$\mathcal{B}(f_2(1270) \to K^{*}K )$       &    $0.273\pm0.048$       & $\mathcal{B}(f_2'(1525) \to K^{*}K )$     &    $22.28\pm2.80$          \\\hline
\end{tabular} \label{Tab:NunmT2VP}
\end{table}

\subsection{Results of the  the $D\to T(T\to VP) \ell^+\nu_{\ell}$ decays }

Using the predictions of   $\mathcal{B}(D \rightarrow T \ell^+\nu_\ell)$ given in Tab. \ref{Tab:BrD2Tlv}  and $\mathcal{B}(T \rightarrow V P)$ given in Tab. \ref{Tab:NunmT2VP},  the  branching ratio predictions of the $D\to V P\ell^+\nu_\ell$ decays with tensor resonances can be obtained, and they are listed in
 the third and fifth columns of  Tabs. \ref{Tab:D2PVlvAT1}-\ref{Tab:D2PVlvAT3}.

As given in Tabs. \ref{Tab:D2PVlvAT1}-\ref{Tab:D2PVlvAT3}, some processes have both tensor and   axial-vector resonant contributions, while other processes only receive axial-vector resonant contributions.  Comparing the tensor resonant contributions with  the  axial-vector resonant contributions in the decays with the same initial states and the same final states, one can see that the axial-vector resonant contributions are dominant.


\section{Summary} \label{sec:Summary}
Four-body semileptonic decays  $D\to  VP\ell^+\nu_\ell$ with axial-vector resonance states and tensor  resonance states were studied using the SU(3) flavor symmetry/breaking approach.
The decay amplitudes (vertex coefficients) of the $D \to M \ell^+ \nu_\ell$ ($M \to VP$) decays were related, and the allowed spaces of relevant non-perturbative parameters  of $D \to A \ell^+ \nu_\ell $, $A \to VP$, and $T \to VP$ decays were  obtained  by using the present experimental  data. The form factors of $D \to T\ell^+\nu_\ell$ decays were taken from  the results in the light-front quark model since no experimental data are available at present. Finally, in terms of the predicted branching ratios  of the $D \to M \ell^+ \nu_\ell $ and $M \to VP$ decays,  the branching ratios of the $D\to M(M\to VP)\ell^+\nu_\ell$ decays were  obtained using the narrow width approximation.
In addition, the sensitivity of the branching ratios of the $D \to A \ell^+ \nu_\ell $ and $A \to VP$ decays  to the constrained non-perturbative  parameter spaces was explored.
Our main results can be summarized as follows.

{\bf  1) Decays $D\to A(A\to VP)\ell^+\nu_\ell$}:
$\mathcal{B}(D\to K_1(1270)\ell^+\nu_\ell)$ and $\mathcal{B}(D_s\to f_1/h_1\ell^+\nu_\ell)$ are
 on the order of $\mathcal{O}(10^{-4}-10^{-3})$, which may be measured experimentally in the near future.    $\mathcal{B}(D\to K_1(1400)\ell^+\nu_\ell)$ are strongly suppressed by the mixing angle $\theta_{K_1}$; $\mathcal{B}(D\to  a_1/b_1/f_1/h_1\ell^+\nu_\ell)$ and $\mathcal{B}(D_s\to K_1(1270)/K_1(1400)\ell^+\nu_\ell)$ are strongly suppressed by the CKM matrix element $V_{cd}$,  and  they are on the order of $\mathcal{O}(10^{-6}-10^{-5})$. We also found  that some decay branching ratios are  very sensitive to  the non-perturbative parameters   $A$ and $B$, or the mixing angles $\theta_{K_1}$, $\theta_{3P_1}$, or $\theta_{1P_1}$.
The experimental data for the  $A\to VP$ decays give relatively strong constraints on the non-perturbative  parameters and the mixing angles.
Some branching ratios were obtained with the  large errors  due to  our conservative choice of   parameters and the large experimental errors.
We found that $\mathcal{B}(K_1(1270)\to K^*\pi,K^*\eta)$ are sensitive to  $F$ and $\theta_{K1}$. $\mathcal{B}(K_1(1400)\to \rho K,\omega K,K^*\eta)$  are sensitive to  $D$ and $\theta_{K1}$,
$\mathcal{B}(K_1(1400)\to \rho K,\omega K)$  have  the  inflection point values   when $D\approx-1.0$ and $\theta_{K_1}\approx56^\circ$, and
many $\mathcal{B}(a_1/b_1/f_1/h_1\to VP)$ are very sensitive to  $F$, $D$,  $\theta_{3_{P1}}$ or $\theta_{1_{P1}}$. Any future measurement of these decays can obviously constrain  the relevant  parameters.
Our predictions of  $\mathcal{B}(D^+\to b_1(1235)^0e^+\nu_e,~b_1(1235)^0\to \omega\pi^0)$ and $\mathcal{B}(D^0\to b_1(1235)^-e^+\nu_e,~b_1(1235)^-\to \omega\pi^-)$ are much smaller than their present experimental  upper limits. We found that many  $\mathcal{B}(D\to A\ell^+\nu_\ell,A\to VP)$ are on the order of $\mathcal{O}(10^{-5}-10^{-4})$, and they are expected to be measured in the near future.

{\bf 2) Decays $D\to T(T\to VP)\ell^+\nu_\ell$}:
We found that
the predictions of  $\mathcal{B}(D^0\to K^*_2(1430)^-\ell^+\nu_\ell)$, $\mathcal{B}(D^+\to \overline{K}^*_2(1430)^0\ell^+\nu_\ell)$, $\mathcal{B}(D_s^+\to f_2(1270)\ell^+\nu_\ell)$, $\mathcal{B}(D_s^+\to f_2'(1525)\ell^+\nu_\ell)$, and $\mathcal{B}(D^+\to f_2(1270)e^+\nu_e)$
could reach on the order of $10^{-5}$; nevertheless, other branching ratios of the $D\to T\ell^+\nu_\ell$ are obviously suppressed by the CKM matrix element $V^*_{cd}$.
For the $T\to VP$ decays,
$\mathcal{B}(K^*_2(1430)\to \rho K)$, $\mathcal{B}(K^*_2(1430)\to K^*\pi)$,
 $\mathcal{B}(K^*_2(1430)\to \omega K)$,
$\mathcal{B}(a_2(1320)\to \rho\pi)$ and  $\mathcal{B}(f'_2(1525)\to K^*K)$ are on the order of $10^{-2}-10^{-1}$; however,
$\mathcal{B}(K^*_2(1430)\to K^*\eta,\phi K)$, $\mathcal{B}(a_2(1320)\to K^* K)$, and $\mathcal{B}(f_2(1270)\to K^* K)$ are on the order of $10^{-4}-10^{-3}$.
As for the branching ratios of the $D\to T(T\to VP)\ell^+\nu_\ell$ decays, some processes received  both the tensor and axial-vector resonant contributions.  We  found that the axial-vector resonant contributions were dominant, and the tensor  resonant contributions were obviously suppressed.

Although SU(3) flavor symmetry is approximate, it can still provide very useful information about these decays.  According to our rough predictions, some decays of $D \to M \ell^+ \nu_\ell $,  $M \to VP$, and $D\to M(M\to VP)\ell^+\nu_\ell$   might be measured experimentally in the near future, and  they are useful for testing and understanding the non-perturbative effects in the semileptonic decays.  Nevertheless,  some decay branching ratios are very small and could not be measured recently.

\section*{ACKNOWLEDGEMENTS}
We thank Hai-Bo Li, Mao-Zhi Yang  and Xian-Wei Kang for useful discussions. The work was supported by the National Natural Science Foundation of China (No. 12175088 and No. 12365014).

\section*{Appendix}\label{Sec:Append}
 We present the meson multiplets of the SU(3) flavor group  in this section.
Charmed pseudoscalar mesons are composed of a charm quark and a light anti-quark
\begin{eqnarray}
D^i=\Big( D^0(c\bar{u}),~D^+(c\bar{d}),~D^+_s(c\bar{s})\Big).
\end{eqnarray}
Light pseudoscalar mesons and vector mesons  are represented as \cite{He:2018joe}
\begin{eqnarray}
 P&=&\left(\begin{array}{ccc}
\frac{\pi^0}{\sqrt{2}}+\frac{\eta_8}{\sqrt{6}}+\frac{\eta_1}{\sqrt{3}} & \pi^+ & K^+ \\
\pi^- &-\frac{\pi^0}{\sqrt{2}}+\frac{\eta_8}{\sqrt{6}}+\frac{\eta_1}{\sqrt{3}}  & K^0 \\
K^- & \overline{K}^0 &-\frac{2\eta_8}{\sqrt{6}}+\frac{\eta_1}{\sqrt{3}}
\end{array}\right)\,,\\
V&=&\left(\begin{array}{ccc}
\frac{\rho^0}{\sqrt{2}}+\frac{\omega}{\sqrt{2}} & \rho^+ & K^{*+} \\
\rho^- &-\frac{\rho^0}{\sqrt{2}}+\frac{\omega}{\sqrt{2}} & K^{*0} \\
K^{*-} & \overline{K}^{*0} &\phi
\end{array}\right)\,.
\end{eqnarray}
The pseudoscalar mesons $\eta$ and $\eta'$ are the mixtures of $\eta_1=\frac{u\bar{u}+d\bar{d}+s\bar{s}}{\sqrt{3}}$ and $\eta_8=\frac{u\bar{u}+d\bar{d}-2s\bar{s}}{\sqrt{6}}$ with a mixing angle denoted as $\theta_P$ \cite{Bramon:1992kr}
\begin{eqnarray}
\left(\begin{array}{c}
\eta\\
\eta'
\end{array}\right)\,
=
\left(\begin{array}{cc}
\mbox{cos}\theta_P&-\mbox{sin}\theta_P\\
\mbox{sin}\theta_P&\mbox{cos}\theta_P
\end{array}\right)\,\left(\begin{array}{c}
\eta_8\\
\eta_1
\end{array}\right)\,.
\end{eqnarray}
 In our numerical analysis, we will utilize the mixing angle range of $\theta_P=[-20^\circ,-10^\circ]$ from the Particle Data Group (PDG) \cite{PDG2022}.

The multiplets of the light axial-vector mesons are presented in the matrix of $A$ and $B$, where  $A$ represents the axial-vector mesons with $J^{PC}=1^{++}$ and $B$ corresponds to the  mesons with $J^{PC}=1^{+-}$ \cite{Roca:2003uk}
\begin{eqnarray}
A&=&\left(\begin{array}{ccc}
\frac{a^0_1}{\sqrt{2}}+\frac{f_1}{\sqrt{3}}+\frac{f_8}{\sqrt{6}} & a^+_1 & K^+_{1A} \\
a^-_1 &-\frac{a^0_1}{\sqrt{2}}+\frac{f_1}{\sqrt{3}}+\frac{f_8}{\sqrt{6}}  & K^0_{1A} \\
K^-_{1A} & \overline{K}^0_{1A} &\frac{f_1}{\sqrt{3}}-\frac{2f_8}{\sqrt{6}}
\end{array}\right)\,,\\
B&=&\left(\begin{array}{ccc}
\frac{b^0_1}{\sqrt{2}}+\frac{h_1}{\sqrt{3}}+\frac{h_8}{\sqrt{6}} & b^+_1 & K^+_{1B} \\
b^-_1 &-\frac{b^0_1}{\sqrt{2}}+\frac{h_1}{\sqrt{3}}+\frac{h_8}{\sqrt{6}}  & K^0_{1B} \\
K^-_{1B} & \overline{K}^0_{1B} &\frac{h_1}{\sqrt{3}}-\frac{2h_8}{\sqrt{6}}
\end{array}\right)\,.
\end{eqnarray}
Mesons $K_{1}(1270)$ and $K_{1}(1400)$  are the mixtures of $K_{1A}$ and $K_{1B}$  by the mixing angle $\theta_{K_1}$
\begin{eqnarray}
\left(\begin{array}{c}
K_1(1270)\\
K_1(1400)
\end{array}\right)\,
=
\left(\begin{array}{cc}
sin\theta_{K_1}&cos\theta_{K_1}\\
cos\theta_{K_1}&-sin\theta_{K_1}
\end{array}\right)\,
\left(\begin{array}{c}
K_{1A}\\
K_{1B}
\end{array}\right).
\end{eqnarray}\label{Eq:etamix}
Mesons  $f_1(1285)$, $f_1(1420)$  ($h_1(1170)$, $h_1(1415)$) can be expressed by the mixing angle  $\theta_{3P_1}$ ($\theta_{1P_1}$)
\begin{eqnarray}
\left(\begin{array}{c}
f_1(1285)\\
f_1(1420)
\end{array}\right)\,
=
\left(\begin{array}{cc}
cos\theta_{3P_1}&sin\theta_{3P_1}\\
-sin\theta_{3P_1}&cos\theta_{3P_1}
\end{array}\right)\,
\left(\begin{array}{c}
f_{1}\\
f_{8}
\end{array}\right),\label{Eq:etamix}\\
\left(\begin{array}{c}
h_1(1170)\\
h_1(1415)
\end{array}\right)\,
=
\left(\begin{array}{cc}
cos\theta_{1P_1}&sin\theta_{1P_1}\\
-sin\theta_{1P_1}&cos\theta_{1P_1}
\end{array}\right)\,
\left(\begin{array}{c}
h_{1}\\
h_{8}
\end{array}\right).
\end{eqnarray}\label{Eq:etamix}
The value of mixing angle $\theta_{K_1}$ has been investigated in various studies, for examples,   $33^\circ$ and $57^\circ$ in Ref. \cite{Suzuki:1993yc}, $ 37^\circ$ or $58^\circ$ \cite{Cheng:2003bn},  [$35^\circ,55^\circ$] \cite{Burakovsky:1997dd},  $60^\circ$ \cite{Tayduganov:2011ui}, $(33.6\pm4.3)^\circ$ \cite{Divotgey:2013jba}, etc.
These ranges will not be used in later numerical analysis. We will obtain the ranges of $\theta_{K_1}$, $\theta_{1P_1}$ and $\theta_{3P_1}$ from  the $D \to A(A \to VP)\ell^+\nu_\ell$ decays.

The lowest multiplet of p-wave tensor mesons with $J^{PC}=2^{++}$ take the form as \cite{Ecker:2007us,Chen:2023ybr}
\begin{eqnarray}
 T&=&\left(\begin{array}{ccc}
\frac{a_2^0}{\sqrt{2}}+\frac{f_2^q}{\sqrt{2}} & a^+_2 & K_2^{*+}  \\
a^-_2 & -\frac{a^0_2}{\sqrt{2}}+\frac{f^q_2}{\sqrt{2}} & K^{*0}_2 \\
K^{*-}_2 & \overline{K}^{*0}_2 & f^s_2
\end{array}\right)\,.
\end{eqnarray}
$f_{2}(1270)$ and $f_2'(1525)$ are mixed by $f_2^q$ and $f_2^s$ with the mixing angle $\theta_{f_2}$
\begin{eqnarray}
\left(\begin{array}{c}
f_2(1270)\\
f_2'(1525)
\end{array}\right)\,
=
\left(\begin{array}{cc}
cos\theta_{f_2}&sin\theta_{f_2}\\
sin\theta_{f_2}&-cos\theta_{f_2}
\end{array}\right)\,
\left(\begin{array}{c}
f^q_2\\
f^s_2
\end{array}\right),
\end{eqnarray}
where $~\theta_{f_2}\in[8^\circ, 10^\circ]$ \cite{Cheng:2010yd} will be used in our calculation.

\end{document}